# Carbon Cycle Imbalances on Arid Terrestrial Planets with Implications for Venus

Haskelle T. White-Gianella[1] and Joshua Krissansen-Totton[1,2]
[1] Department of Earth and Space Sciences/Astrobiology Program, University of Washington, Seattle, WA 98195, USA; hasktw@uw.edu
[2] NASA NExSS Virtual Planetary Laboratory, University of Washington, Seattle, WA 98195, USA

## Abstract

Arid terrestrial exoplanets are potentially abundant and are thus interesting targets in the search for life. In particular, M-dwarf planets such as those in the TRAPPIST-1 system may possess limited surface water, whereas early solar system terrestrials may have had small surface water inventories postmagma ocean solidification. On modern Earth, there is enough surface water for a balanced geologic carbon cycle, meaning silicate weathering balances the volcanic outgassing of $CO_2$. However, on arid planets, there may not be enough surface water for this silicate weathering thermostat to maintain habitable conditions. Here, we show that arid planets enter a regime where weathering cannot keep up with volcanic degassing of $CO_2$. Using a coupled model of the geologic carbon cycle, we find that terrestrial Earth-like planets require an initial surface water inventory of at least ∼20%–50% of Earth's ocean mass to maintain a balanced geologic carbon cycle and temperate surface temperature over 4.5 Gyr of evolution. Arid planets with less than ∼20%–50% of Earth's oceans cannot maintain high silicate weathering fluxes, potentially causing a runaway increase in atmospheric $CO_2$. In addition, we explore Venus-like instellations and find that limited surface water could have destabilized Venus's carbon cycle, triggering a transition from temperate to uninhabitable. Even if a planet resides in the habitable zone of its star, if arid, it may transition to an uninhabitable state due to an unbalanced carbon cycle. More broadly, arid terrestrial exoplanets are less likely to remain habitable on long timescales, and may thus be poor candidates for biosignature searches.

*Unified Astronomy Thesaurus concepts:* Planetary science (1255); Venus (1763); Extrasolar rocky planets (511); Solar system terrestrial planets (797); Atmospheric evolution (2301)

## 1. Introduction

In the search for life beyond Earth, the presence of liquid water is the primary criterion for assessing habitability. Life requires a liquid solvent to grow, evolve, and sustain metabolism, and liquid water is the only known compound that facilitates the chemical reactions essential for life (A. Pohorille & L. R. Pratt 2012; C. Cockell et al. 2016). The long-term stability of surface liquid water depends on a delicate balance of temperature, pressure, and chemical composition of a planet. These factors, in turn, are shaped by planetary geophysical evolution and astronomical influences such as stellar spectral energy distribution, stellar age, and orbital dynamics (V. S. Meadows & R. K. Barnes 2018).

Stellar characteristics influence the size of the conventional habitable zone (HZ): the region around a star where liquid water can exist on the surface of orbiting planets with $N_2$–$CO_2$–$H_2O$ atmospheres (J. F. Kasting et al. 1993; R. K. Kopparapu et al. 2013). The inner edge of the HZ is defined by the runaway greenhouse limit, which is an incoming shortwave radiation threshold beyond which a planet's outgoing longwave radiation (OLR) no longer increases with surface temperature, causing uncontrolled heating. This occurs because rising temperatures enhance water vapor concentrations to the point where the atmosphere becomes optically thick in the infrared, trapping heat and preventing radiative cooling. Once this limit is crossed, it may trigger the complete evaporation of surface water to restore a radiative balance (A. P. Ingersoll 1969; J. B. Pollack 1971; J. F. Kasting 1988). Planets too close to their star cannot sustain liquid water for this reason, defining the inner edge of the HZ. Conversely, the outer edge of the HZ is defined by the maximum $CO_2$ greenhouse limit, where $CO_2$ concentrations are high enough that cooling from Rayleigh scattering dominates over the warming greenhouse effect (R. K. Kopparapu et al. 2013). The enhanced cooling effect ultimately limits surface liquid water. Both the inner and outer edges are highly dependent on albedo, and their position can dramatically change depending on the cloud dynamics of the planet (M. J. Way et al. 2016; D. Kitzmann 2017).

However, whether a planet within the HZ can actually support continuous habitability is also dependent on atmospheric composition. Even if a planet is within the conventional HZ, elevated surface temperatures can weaken the atmospheric cold trap, allowing water vapor to accumulate in the upper atmosphere. This so-called moist greenhouse state can accelerate water loss via water vapor photolysis and hydrogen escape (J. F. Kasting 1988). A planet near the outer edge may receive enough incoming radiation, but if its atmosphere has no $CO_2$, all surface water would freeze (J. F. Kasting et al. 1993). Indeed, by many estimates, modern Mars resides within the Sun's HZ but does not sustain liquid surface water because of the lack of sufficient greenhouse warming (G. M. Martínez et al. 2017). Thus, a terrestrial planet in the HZ needs the right atmospheric composition and albedo to sustain surface temperatures suitable for liquid water. For planets like Earth with abundant surface water, atmospheric $CO_2$ within the HZ is assumed to be regulated by the carbonate–silicate weathering thermostat, as discussed below. This defines the "geochemical HZ" where geochemical cycling adjusts $CO_2$ concentrations across the HZ to maintain habitable conditions (R. J. Graham &

R. T. Pierrehumbert 2024). Even the presence of liquid water does not guarantee habitability. Liquid water might exist on a terrestrial planet with a 100 bar $CO_2$ atmosphere and 450 K surface, but the temperatures would be too hot for biochemistry since they exceed the thermal limit of cell division and protein denaturing (A. Clarke 2014). In this study, we define habitability more conservatively as the temperature range on terrestrial planets suitable for Earth-like life, which is approximately −20°C (−253 K) to 122°C (395 K) (A. Clarke 2014).

Arid planets with low surface water inventories (<<1 Earth ocean) may be a common outcome of planetary evolution (S. Seager & D. Deming 2010; G. D. Mulders et al. 2021; S. Sabotta et al. 2021). Terrestrial planets in M-dwarf systems typically form from less massive protoplanetary disks and may form with small water inventories (I. Baraffe et al. 1998; I. Baraffe et al. 2015; R. Luger & R. Barnes 2015). Even if these planets do not form with low water inventories, M-dwarf terrestrials might become arid due to extensive water loss during their star's superluminous pre-main sequence (R. M. Ramirez & L. Kaltenegger 2014; R. Wordsworth & R. Pierrehumbert 2014; R. Luger & R. Barnes 2015; E. Bolmont et al. 2017; J. Krissansen-Totton et al. 2024). Given arid starting conditions, it is unknown how HZ M-dwarf terrestrials would evolve subsequently after the pre-main sequence. Moreover, even terrestrial planets around Sun-like stars may form arid. During magma ocean solidification, rapid movement of the solidification front traps highly soluble $H_2O$ in the mantle, but releases less soluble $CO_2$ into the atmosphere (S. Hier-Majumder & M. M. Hirschmann 2017; D. J. Bower et al. 2022; J. Krissansen-Totton & J. J. Fortney 2022; Y. Miyazaki & J. Korenaga 2022). Given that rocky planets may often form with dry, $CO_2$-rich atmospheres, it is worth considering how planets evolve given these arid initial conditions.

While they may be common outcomes of planet formation, the habitability prospects for arid terrestrial planets are unknown. F. Ding & R. D. Wordsworth (2020) show that the atmospheric hydrological cycle may be stable on arid, tidally locked terrestrial planets, depending on the strength of nightside cold traps, surface topography, planetary rotation rate, and the global weathering rate. Water transport between the substellar tropopause and the nightside cold traps can also sustain the water cycle under arid conditions (F. Ding & R. D. Wordsworth 2021). General circulation models (GCMs) from Y. Abe et al. (2011) show that dry planets are less prone to a runaway greenhouse phase compared to waterworlds (planets with >>1 Earth ocean) due to reduced atmospheric humidity and limited water vapor greenhouse forcing. The spatial distribution of surface water and continental configuration further influence short-term climate stability (T. Kodama et al. 2018, 2019). In particular, whether continents cluster around the poles versus the equator can warm the global mean surface temperature on arid surfaces by 2°C–5°C (D. M. Glaser et al. 2025). However, the long-term climate stability of a planet also depends on the geologic carbon cycle. This crucial geochemical feedback has been shown to destabilize at the outer portions of the HZ (R. J. Graham & R. T. Pierrehumbert 2024), but whether it can remain balanced on arid terrestrial planets throughout the HZ is uncertain.

### 1.1. The Geologic Carbon Cycle

The geologic carbon cycle describes the exchange of carbon between the surface and the interior. This cycle is a key factor in sustaining long-term habitability due to its role in regulating atmospheric $CO_2$ and surface temperatures (J. C. G. Walker et al. 1981; R. A. Berner et al. 1983). On planets with liquid surface water, atmospheric $CO_2$ dissolves in water to form carbonic acid, which rains on the surface and chemically weathers continental silicate rocks, releasing cations such as $Ca^{2+}$ and $Mg^{2+}$. These ions are carried to the ocean by runoff, where carbonate rocks precipitate, removing carbon from the atmosphere–ocean system. The temperature dependence of weathering establishes a negative feedback that stabilizes surface climate over geologic timescales: higher atmospheric $CO_2$ increases temperatures, which enhances rock weathering, which draws down more $CO_2$, thereby stabilizing climate against external forcings (J. C. G. Walker et al. 1981; E. Tajika & T. Matsui 1992; R. A. Berner & K. Caldeira 1997; N. H. Sleep & K. Zahnle 2001). The weathering of silicates and deposition of carbonates may also occur in the seafloor as carbon-bearing seawater circulates through oceanic crust (L. A. Coogan & K. M. Gillis 2013; J. Krissansen-Totton & D. C. Catling 2017). On planets with plate tectonics, subduction zones recycle these carbonates into the mantle, where they may eventually be returned to the atmosphere via metamorphic or magmatic processes (R. A. Berner 2004). While plate tectonics is not strictly required for a functioning carbon cycle (B. J. Foley & A. J. Smye 2018), it enhances the supply of fresh, weatherable rock that is critical for $CO_2$ drawdown (B. J. Foley 2015), and provides a mechanism to remove $CO_2$ from the atmosphere into the mantle via subduction (B. J. Foley & A. J. Smye 2018).

On Earth, the surface water inventory is sufficient to permit high continental weathering fluxes, thereby maintaining habitable conditions via this silicate weathering thermostat (J. C. G. Walker et al. 1981). In other words, there is enough precipitation and runoff to balance the input of $CO_2$ into the atmosphere from volcanic outgassing (and consequent silicate weathering and carbonate deposition). The ability of silicate weathering to regulate climate is thus dependent on several hydrologic and tectonic processes that affect runoff (K. Maher & C. P. Chamberlain 2014). R. J. Graham & R. Pierrehumbert (2020) were the first to quantify these hydrologic factors and their effects on planetary habitability on ocean-bearing terrestrial planets.

In the outer HZ, R. J. Graham & R. T. Pierrehumbert (2024) found that silicate weathering fails as instellation decreases, and the partial pressure of $CO_2$ ($pCO_2$) increases, narrowing the width of the geochemical HZ relative to the conventional HZ. Within the HZ, the carbon cycle dynamics of arid planets are largely underexplored. F. Ding & R. D. Wordsworth (2020) adopted a temperature and $pCO_2$-dependent model of silicate weathering to argue that there is a critical $CO_2$ outgassing threshold needed to maintain dayside liquid water on tidally locked arid planets. However, shallow oceans and meager hydrological cycles on arid planets might limit continental silicate weathering, which jeopardizes the balance with volcanic outgassing of $CO_2$ (Y. Abe et al. 2011). Small water inventories may thus limit the ability of the geologic carbon cycle to maintain a habitable climate.

### 1.2. Applications to Venus

Whether the geologic carbon cycle can operate on arid planets is especially relevant for early Venus. The modern surface is uninhabitable, with average temperatures of 460°C, a surface pressure 92 times that of Earth, and a dense $CO_2$-dominated atmosphere (S. W. Bougher et al. 1997). However, Venus could have been habitable in the past under the faint young Sun (J. B. Pollack 1971; M. J. Way et al. 2016). The uncertainties of Venus's past cloud-albedo feedback (M. J. Way et al. 2016) and atmospheric evolution (J. Krissansen-Totton et al. 2021a) place it on the edge of the inner HZ, where the timing and trigger of its runaway greenhouse state are unclear (K. Hamano et al. 2013). There is even some tentative evidence to suggest that Venus was habitable in the past. GCM's from M. J. Way et al. (2016) demonstrate that, with sufficient dayside cloud cover paired with its slow rotation rates, Venus's surface could have maintained habitable surface temperatures up to 715 million years ago. Physical evidence supporting a temperate past climate includes possible remnants of felsic continental crust that typically form in the presence of water (M. S. Gilmore et al. 2015; P. G. Resor et al. 2021) and D/H ratios indicative of larger—although not necessarily condensed—surface water reservoirs in the past (T. M. Donahue et al. 1982; T. Donahue 1999). Contrasting this is evidence supporting a scenario where Venus was always in a runaway greenhouse since formation. Crustal plateaus on Venus are hypothesized to be felsic, but plateaus composed of felsic minerals would likely collapse from the viscous flowing lower crust (F. Nimmo & S. Mackwell 2023). Furthermore, the atmosphere–cloud dynamics postaccretion may have precluded surface liquid water condensation (M. Turbet et al. 2021).

While Venus's past climate is still uncertain, if it was habitable in the past, then it must have experienced a significant climate transition to achieve its modern conditions. One explanation for Venus's modern inhospitable state is that, as the Sun's luminosity increased, the planet received more incoming radiation, warming the previously habitable surface, eventually triggering a transition to a runaway greenhouse state (C. Goldblatt et al. 2013). In this scenario, any surface water would have evaporated into steam, where photodissociation of water led to rapid hydrogen escape, resulting in the present-day desiccated, $CO_2$-dominated atmosphere (J. F. Kasting 1988). However, increasing luminosity alone cannot explain a climate transition from habitable to uninhabitable. The runaway greenhouse limit is a function of albedo and atmospheric composition (A. Salvador et al. 2017), which is governed by long-term interactions between the atmosphere and interior. If Venus had the albedo conditions shown in M. J. Way et al. (2016) to maintain surface habitability for billions of years, then increasing solar insolation only increases the dayside cloud deck, and thus the albedo. This stabilizing cloud-albedo feedback could have supported temperate surface conditions on Venus even under modern insolation (M. J. Way & A. D. Del Genio 2020).

Another explanation for Venus's current uninhabitable state is that an external forcing other than increasing luminosity terminated a habitable period. M. J. Way & A. D. Del Genio (2020) propose that habitable conditions could have even persisted to the present day were it not for catastrophic volcanic activity through episodic or quasi-continuous global resurfacing (R. G. Strom et al. 1994) or the emplacement of large igneous provinces (M. J. Way et al. 2022), any of which could have acted as potential triggers. However, changes in outgassing alone may not be sufficient to cause a permanent change in the climate state. If Venus once had surface liquid water, silicate weathering could have acted to regulate atmospheric $CO_2$ concentrations (J. F. Kasting et al. 1984), even in the presence of rapid volcanic outgassing. Additionally, large igneous province eruptions expose fresh basaltic rock to the surface, significantly enhancing $CO_2$ drawdown through chemical weathering (L. Johansson et al. 2018). It is unlikely that changes to outgassing alone can permanently destabilize a climate with a working carbonate–silicate feedback; limits to weathering are needed as well. Stagnant-lid decarbonation feedback has also been a proposed explanation for Venus's terminated habitability (D. Höning et al. 2021). However, it is unknown if early Venus was in a stagnant-lid tectonic regime, and there is some evidence for localized subduction (C. Gillmann & P. Tackley 2014; A. Davaille et al. 2017; L. S. Zampa et al. 2018).

Here, we explore an alternative explanation for Venus's modern postrunaway greenhouse state. Venus may have started with small surface water inventories, which is supported by magma ocean solidification models (S. Hier-Majumder & M. M. Hirschmann 2017; D. J. Bower et al. 2022; Y. Miyazaki & J. Korenaga 2022), planet formation models demonstrating stochastic variations in the initial water inventories of solar system terrestrials (S. N. Raymond et al. 2004), and the fact that Venus presumably formed in a hotter, more volatile-poor region of the protoplanetary disk than Earth. Under this initial arid regime, the lack of precipitation may have limited silicate weathering fluxes and $CO_2$ drawdown. Without a working silicate weathering thermostat to balance outgassing, $CO_2$ may have accumulated in the atmosphere, warming the surface until all surface water was lost, leading to its current uninhabitable state.

### 1.3. This Paper in Context

In this work, we apply a geologic carbon cycle model to terrestrial planets with low surface water inventories to explore their long-term habitability prospects. In Section 2, we describe the main components of the carbon cycle model, which builds on past atmosphere–interior evolution models (Section 2.1) that include parameterizations for planetary climate, atmospheric escape (Sections 2.1.1, 2.1.2; J. Krissansen-Totton et al. 2018, 2021b), and a global weathering model adapted from R. J. Graham & R. Pierrehumbert (2020; Section 2.2). New additions to the model include wind-driven evaporation scaled to the land fraction (Section 2.2.3), an updated outgassing formulation (Section 2.3), and three end-member parameterizations for the deep-water cycle (Section 2.4). Model results are shown in Section 3, and their implications for long-term habitability are discussed in Sections 4 and 5.

### 2. Methods

We model the evolution of the geologic carbon cycle by tracking the fluxes of water and carbon between the interior and the atmosphere–ocean system. Our model, loosely based on J. Krissansen-Totton et al. (2021b), couples outgassing/weathering, stellar evolution, planetary hypsometry, and the deep-water cycle. Here, we summarize the main features of the model described in Figure 1. The Python model code is available on Zenodo via doi:10.5281/zenodo.18706908.

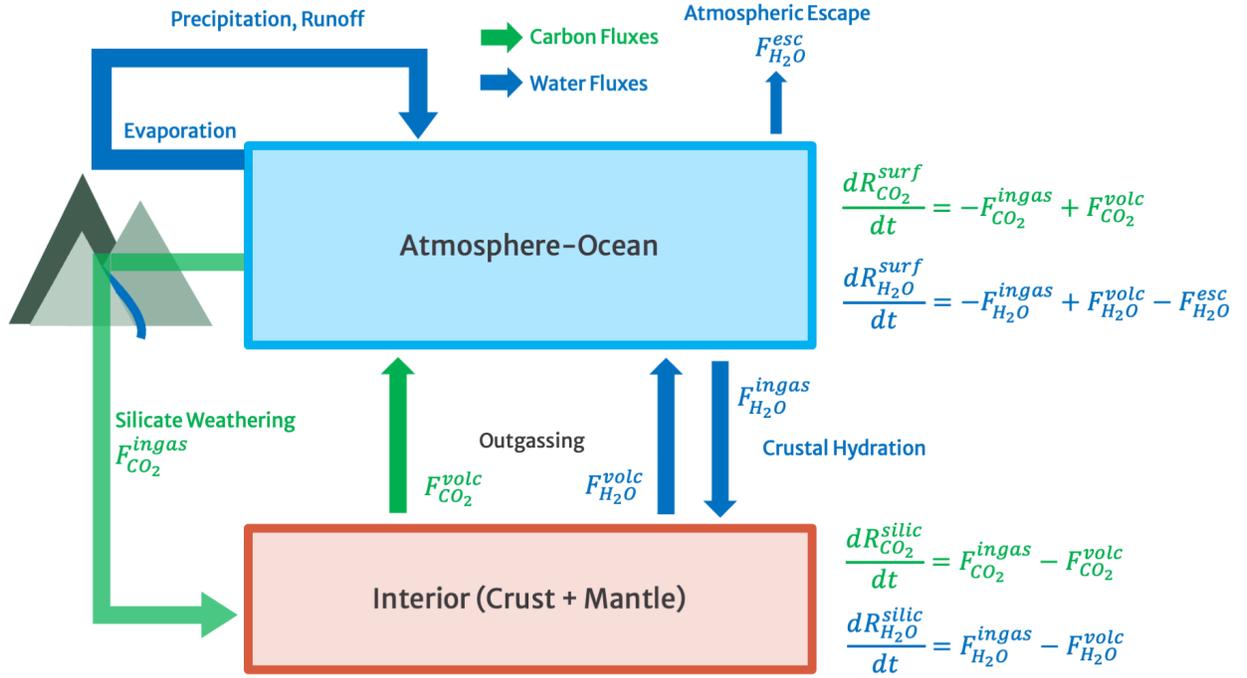

**Figure 1.** Schematic representation of the box model used in this study. Carbon fluxes are denoted by green arrows, and water fluxes are denoted by blue arrows. Our model is divided into two boxes: the surface (atmosphere and ocean) and the interior (mantle and crust). Equations (1)–(4) govern the transfer of H$_2$O and CO$_2$ in the system.

### 2.1. Atmosphere–Interior Evolution Model

We model the exchange of water between the interior and atmosphere of a planet with a system of ordinary differential equations. The change in the interior water reservoir with time is represented by

$$\frac{dR_{H_2O}^{silic}}{dt} = F_{H_2O}^{ingas} - F_{H_2O}^{volc} \quad (1)$$

where $R_{H_2O}^{silic}$ [kg] is the reservoir of water stored in the mantle and crust, $F_{H_2O}^{ingas}$ [kg s$^{-1}$] is the water ingassing flux, and $F_{H_2O}^{volc}$ [kg s$^{-1}$] is the outgassing flux of water. The ingassing flux represents any process that removes water from the surface and stores it in silicates, including subduction, crustal hydration, etc. Ingassing adds water into the crust or mantle, while volcanic outgassing transfers water from the interior to the atmosphere. We do not distinguish between crustal and mantle reservoirs for simplicity, although we do explore various end-member ingassing cases to represent different tectonic regimes (see below). Note, however, that, without explicitly modeling subduction, hydrated crust may not be recycled into the mantle to feed partial melting and mantle degassing.

The change in the surface water reservoir with time is described by

$$\frac{dR_{H_2O}^{surf}}{dt} = -F_{H_2O}^{ingas} + F_{H_2O}^{volc} - F_{H_2O}^{esc} \quad (2)$$

where $R_{H_2O}^{surf}$ [kg] is the reservoir of surface water (oceans, ice, and water vapor), and a new term, $F_{H_2O}^{esc}$ [kg s$^{-1}$], describes the loss of water to space via diffusion-limited escape.

The exchange of carbon is represented by similar equations:

$$\frac{dR_{CO_2}^{silic}}{dt} = F_{CO_2}^{ingas} - F_{CO_2}^{volc} \quad (3)$$

$$\frac{dR_{CO_2}^{surf}}{dt} = -F_{CO_2}^{ingas} + F_{CO_2}^{volc}. \quad (4)$$

$R_{CO_2}^{silic}$ [kg] represents the interior reservoir of carbon stored in the crust and mantle, and $R_{CO_2}^{surf}$ [kg] is all carbon stored in the atmosphere and ocean. $F_{CO_2}^{ingas}$ [kg s$^{-1}$] is the ingassing flux of carbon dioxide, which represents carbonate formation due to silicate weathering. $F_{CO_2}^{volc}$ [kg s$^{-1}$] is the outgassing flux of carbon dioxide representing the degassing of CO$_2$ from volcanoes. Equation (3) describes the change in interior carbon with time, while Equation (4) shows the change in atmospheric carbon (we assume CO$_2$ escape is negligible).

#### 2.1.1. Climate Model

We use a full radiative–convective climate model for CO$_2$–H$_2$O–N$_2$ atmospheres as outlined in J. Krissansen-Totton et al. (2021b) to self-consistently calculate surface temperature and the atmosphere–surface partitioning of water. We calculated OLR using the DISORT-based radiative transfer code from E. Marcq et al. (2017), which includes H$_2$O and CO$_2$ opacities (S. A. Clough et al. 2005; B. Bézard et al. 2011) determined by our model's surface reservoir of water and carbon dioxide. The atmospheric structure is assumed to follow a dry adiabat (when water is supercritical or subsaturated), then a moist adiabat, and then an isothermal upper layer (J. F. Kasting 1988); the temperature–pressure profile also sets the water vapor abundance profile, and we assume 100% relative humidity (this will maximize escape fluxes, but later, we consider a no-escape end-member). Absorbed shortwave radiation (ASR) is calculated using the globally averaged stellar flux scaled by assumed Bond albedo. Albedo is fixed in our calculations since cloud feedback cannot be realistically modeled with a 1D radiative–convective climate model. However, to explore composition- and cloud-dependent albedo effects, we present sensitivity tests with widely varying fixed albedos. Given surface inventories of $H_2O$ and $CO_2$, at each time step, we solve for the surface temperature that ensures OLR balances ASR using Python's *solve_ivp* solver (P. Virtanen et al. 2020). OLR is also a function of dissolved carbonate concentrations in the ocean, and we assume a constant carbonate saturation state following E. W. Schwieterman et al. (2019). For computational efficiency, we precomputed a grid of OLR values as a function of surface temperature (250–4000 K), surface water (10 Pa–1 GPa), surface carbon dioxide (10 Pa–0.1 GPa), and planetary skin temperature (150–350 K). Within the grid, we use linear interpolation between grid points. In the rare cases where the model extends beyond the grid, we apply linear extrapolation. A 1 bar partial pressure of $N_2$ background is assumed at every grid point. The resulting stratospheric $H_2O$ mixing ratios are used to determine atmospheric escape rates.

### 2.1.2. Atmospheric Escape

The reservoir of surface water is controlled, in part, by atmospheric escape. We calculate atmospheric escape of water assuming diffusion-limited escape of hydrogen through a background atmosphere composed of $CO_2$ and $N_2$, following R. D. Wordsworth & R. T. Pierrehumbert (2013b):

$$F_{H_2O}^{esc} = b_{H_2O} f_{H_2O} \left( \frac{1}{H_n} - \frac{1}{H_{H_2O}} \right) (4\pi r^2 M_{H_2O}). \quad (5)$$

$F_{H_2O}^{esc}$ [kg s$^{-1}$] is the diffusion-limited escape flux. $b_{H_2O}$ [mol m$^{-1}$ s$^{-1}$] is the effective binary diffusion coefficient of the gas species involved (K. J. Zahnle & J. F. Kasting 1986), $f_{H_2O}$ is the stratospheric water mixing ratio, $H_n$ [m] is the scale height of the background gases, and $H_{H_2O}$ [m] is the scale height of water. The escape flux is then converted from mol $H_2O$ m$^{-2}$ s$^{-1}$ to kg s$^{-1}$ by applying it over the surface area of the planet ($4\pi r^2$ [m$^2$]) multiplied by the molar mass of water ($M_{H_2O}$ = 0.018 kg mol$^{-1}$). The diffusion coefficient and scale heights depend on the temperature, composition, and mixing ratios of the stratosphere derived from the climate model. The escape flux $F_{H_2O}^{esc}$ is thus determined by the water vapor mixing ratio in the upper atmosphere and the temperature-dependent diffusion of $H_2O$ through noncondensable gases ($CO_2$, $O_2$, and $N_2$), which are weighted by their atmospheric abundances. Escape is limited to hydrogen species, which is reasonable for the long-term evolution of temperate, Earth-sized planets around G-type stars. We ignore $O_2$ generated by H escape and implicitly assume it is consumed by crustal sinks, as has occurred on Venus (C. Gillmann et al. 2009). Evolutionary models estimate that crustal oxidation could have removed tens to hundreds of bars of $O_2$, roughly equivalent to 4%–100% of Earth's oceans (C. Gillmann et al. 2009; J. Krissansen-Totton et al. 2021a; A. O. Warren & E. S. Kite 2023). Therefore, for the small water inventories considered here (0.1%–100% of Earth's oceans), $O_2$ may have been removed to crustal sinks before it could have a significant impact on climate and carbon cycle feedback. We explore carbon cycle evolution both with and without atmospheric escape to show that a carbon cycle imbalance on arid planets is not primarily a consequence of hydrogen escape.

### 2.2. Weathering Models

#### 2.2.1. WHAK

Our weathering parameterization controls $F_{CO_2}^{ingas}$ in our atmosphere–interior evolution model and builds on previous geologic carbon cycle models. J. C. G. Walker et al. (1981) was the first to propose a negative feedback cycle used to modulate atmospheric $CO_2$ and stabilize the Earth's climate. Their original weathering model, the Walker, Hays, and Kasting (WHAK) model, was dependent only on the $pCO_2$ in the atmosphere and temperature:

$$W/W_0 = (P/P_0)^{0.3} \exp\left( \frac{\Delta T}{13.7} \right) \quad (6)$$

where $W_0$ [mol yr$^{-1}$] is the modern silicate weathering flux, $P_0$ [ppm] is the present-day $pCO_2$, $P$ [ppm] is the atmospheric concentration of $CO_2$, and $\Delta T$ [K] is the difference in temperature from the present-day value of 288 K. WHAK is based on experiments constraining feldspar dissolution rates at different $pCO_2$, which is how the 0.3 power dependency was attained (M. Lagache 1976). 3D climate models from R. T. Wetherald & S. Manabe (1975) provided the basis of the temperature-dependent runoff constant (13.7 K). The WHAK model provided the fundamental framework for geologic carbon cycle modeling. However, WHAK's control on atmospheric $CO_2$ is primarily dependent on the kinetics of silicate dissolution and does not account for several critical processes.

#### 2.2.2. MAC

K. Maher & C. P. Chamberlain (2014) developed WHAK by including concentration-dependent thermodynamic limits to weathering rates. The Maher–Chamberlain (MAC) formulation also added an explicit runoff dependence to weathering rates, so that weathering can be kinetically, thermodynamically, or runoff limited depending on input parameters. R. J. Graham & R. Pierrehumbert (2020) were the first to apply MAC to exoplanetary conditions and highlight the importance of using MAC over WHAK to assess planetary habitability. Their weathering model, which includes limits to precipitation and runoff, is the basis of the weathering model used here. MAC relates solute transport as a function of mean fluid travel time in a drainage basin to the soil age of the weathering front. Weathering-derived solute is calculated as a function of the dimensionless Damköhler number ($Da$):

$$Da = \frac{R_n L \phi}{q C_{eq}} = \frac{t_f}{t_{eq}}. \quad (7)$$

This equation relates the mean fluid travel time, $t_f \approx \frac{L\phi}{q}$, to the time required for weathering reactions to reach equilibrium, $t_{eq} \approx \frac{C_{eq}}{R_n}$. Runoff is denoted as $q$ [m yr$^{-1}$], $L$ [m] is the flowpath length, $\phi$ is the effective porosity, and $R_n$ [$\mu$mol l$^{-1}$ yr$^{-1}$] is the reaction rate for $n$ mineral composition.

Table 1
Fixed Variables Used in MAC Weathering Model

| Variable | Name/Description | Units | Value from R. J. Graham & R. Pierrehumbert (2020) |
| --- | --- | --- | --- |
| $\gamma$ | Land fraction | ⋯ | Computed from climate model |
| $pCO_2$ | Atmospheric $CO_2$ pressure | bar | " |
| $T$ | Surface temperature | K | " |
| $q$ | runoff | m yr$^{-1}$ | Computed from precipitation model |
| $\tau$ | Scaling constant | ⋯ | $e^2 \approx 7.39$ |
| $\Lambda$ | equilibrium coefficient | ⋯ | 0.0014 |
| $T_{\text{ref}}$ | Modern global average reference temperature | K | 288.0 |
| $pCO_{2,\text{ref}}$ | Preindustrial partial pressure of $CO_2$ | bar | $2.8 \times 10^{-4}$ |
| $n$ | Thermodynamic $pCO_2$ dependence | ⋯ | 0.316 |
| $\beta$ | $CO_2$ dependence kinetic weathering | ⋯ | 0.2 |
| $L$ | Flow path length | m | 1.0 |
| $\phi$ | Porosity | ⋯ | 0.1 |
| $\rho_{sf}$ | Mass of mineral to fluid volume ratio | g l$^{-1}$ | 12,728.0 |
| $A$ | Specific surface area | m$^2$ kg$^{-1}$ | 100.0 |
| $X_r$ | Mineral concentration in fresh rock | $g_{\text{min}}/g_{\text{soil}}$ | 0.36 |
| $k_{\text{eff,ref}}$ | Reference net weathering rate constant | mol m$^{-2}$ s$^{-1}$ | $8.7 \times 10^{-6}$ |
| $m$ | Molar mass of the weathering minerals | kg mol$^{-1}$ | 0.27 |
| $\Gamma$ | Fraction of precipitation converted to runoff | ⋯ | 0.2 |

$C_{\text{eq}}$ [$\mu$mol l$^{-1}$] is the maximum solute concentration assuming thermodynamic equilibrium between dissolving and precipitating minerals, i.e., the novel "thermodynamic limit" of the MAC model. Factoring out the runoff $q$ yields the Damköhler coefficient $D_W$ [m yr$^{-1}$]:

$$D_W = \frac{L\phi}{t_{\text{eq}}} = \frac{L\phi R_{n,\max} f_w}{C_{\text{eq}}}. \quad (8)$$

The weathering model now accounts for the fraction of fresh minerals via erosion in the weathering front, $f_w$, which is multiplied with the maximum mineral reaction rate $R_{n,\max}$ [$\mu$mol l$^{-1}$ yr$^{-1}$]. The fresh mineral fraction is equal to the mineral concentration in soil, $X_s$, divided by the mineral concentration in fresh rock ($X_r$):

$$f_w = \frac{X_s}{X_r} = \frac{1}{1 + mk_{\text{eff}} A t_s}. \quad (9)$$

The fraction of fresh minerals, $f_w$ [g g$^{-1}$], is the inverse of the soil age $t_s$ [yr], multiplied by the molar mass $m$ [g mol$^{-1}$], the specific surface area $A$ [m$^2$ g$^{-1}$] of the weathering minerals, and the weathering rate constant $k_{\text{eff}}$ [mol m$^{-2}$ yr$^{-1}$].

To calculate solute production from weathering, $C$ [$\mu$mol l$^{-1}$], Equation (8) is inserted into a series of convolution integrals described in the supplementary material for K. Maher & C. P. Chamberlain (2014). The resulting $C$ is in terms of the Damköhler coefficient (accounting for fresh minerals) and mean fluid travel time:

$$C = C_{\text{eq}} \cdot \frac{\tau D_W/q}{1 + \tau D_W/q}. \quad (10)$$

$\tau$ is a constant equal to $e^2$. Now, solutes produced from weathering are directly related to soil age and runoff in hydrological systems.

Our model uses the MAC formulation described in R. J. Graham & R. Pierrehumbert (2020) to calculate the continental weathering flux as a function of land fraction. Given the surface $H_2O$ and $CO_2$ reservoir, the climate model (Section 2.1.1) calculates the surface temperature and atmosphere–liquid water partitioning. Weathering occurs only when the planet is not submerged in surface oceans, when the surface temperatures are below liquid water's supercritical point (647 K), and if carbon dioxide and liquid water exist on the surface. The maximum equilibrium concentration, $C_{eq}$, is calculated as $C_{eq} = \Lambda \cdot (pCO_2)^n$, which depends on the partial pressure of carbon dioxide ($pCO_2$ [bar]), a weak temperature dependence of the equilibrium constant ($\Lambda = 0.0014$), and a $pCO_2$ dependence ($n$). MAC's weathering rate is computed as a harmonic mean between kinetic and thermodynamic limits. The total weathering flux $w$ [mol m$^{-2}$ yr$^{-1}$] is calculated as $w = C \cdot q$; combining Equations (8)–(10) gives

$$w = \frac{L\phi R_{n,\max} \tau}{\frac{1}{R_n} + mAt_s + \frac{L\phi R_{n,\max}\tau}{qC_{\text{eq}}}}. \quad (11)$$

The weathering rate $w$ is in units of mol m$^{-2}$ yr$^{-1}$. The maximum mineral reaction rate is defined as $R_{n,\max} = \rho_{sf} k_{\text{eff}} A X_r$, where $\rho_{sf}$ [g l$^{-1}$] is the solid mass to fluid volume ratio. Using this definition of $R_{n,\max}$, factoring out common variables, and setting $\alpha = L\phi \rho_{sf} A X_r \tau$ yield

$$w = \frac{\alpha}{\frac{1}{k_{\text{eff}}} + mAt_s + \frac{\alpha}{qC_{\text{eq}}}}. \quad (12)$$

The net weathering rate constant $k_{\text{eff}}$ includes a $pCO_2$ dependence and a temperature dependence on the dissolution rate of silicate minerals. These effects can be described by the WHAK-like Arrhenius relationship as shown in R. J. Graham & R. Pierrehumbert (2020):

$$k_{\text{eff}} = k_{\text{eff, ref}} \exp\left(\frac{T - T_{\text{ref}}}{T_e}\right) \left(\frac{pCO_2}{pCO_{2,\text{ref}}}\right)^\beta \quad (13)$$

where $k_{\text{eff, ref}}$ is the net rate constant at a reference temperature, $T_{\text{ref}}$ is the reference temperature (288 K), $T_e$ is the temperature dependence of kinetic weathering, and $\beta$ is the $pCO_2$ dependence on kinetic weathering. The values used in our MAC weathering model (Equation (12)) are listed in Table 1. For more detailed information on these derivations, refer to

K. Maher & C. P. Chamberlain (2014), M. J. Winnick & K. Maher (2018), R. J. Graham & R. Pierrehumbert (2020), and all accompanying supplementary materials.

This weathering formulation is sensitive to volcanic outgassing, soil age and porosity, land fraction, and runoff. The addition of these dependencies allows the MAC model to assess the carbon cycle and long-term habitability for Earth-like exoplanets within and on the outer edge of the HZ (R. J. Graham & R. T. Pierrehumbert 2024; B. P. Coy et al. 2025). However, to calculate the final weathering flux, $F_{CO_2}^{ingas}$, a new parameterization for precipitation and runoff ($q$ in Equation (11)) is needed for dry terrestrial planets.

### 2.2.3. Precipitation Parameterization

For arid terrestrial planets, the inclusion of runoff in the MAC weathering model can account for the amount of evaporation (which equals precipitation on long timescales) on a planet's surface. For their assessment of the carbonate–silicate thermostat on terrestrial planets in the outer edge of the HZ, R. J. Graham & R. Pierrehumbert (2020) add an energetic limit on precipitation to the MAC framework. Here, maximum planetary precipitation rate is determined by the amount of installation absorbed by surface liquid water, assuming all of this energy is used to evaporate water (latent heat flux). In R. J. Graham & R. Pierrehumbert (2020), the ASR determines the maximum precipitation rate possible, imposing an upper bound on precipitation and runoff; if the actual precipitation rate is below this energetic limit, then precipitation is not determined by ASR and instead scales quasi-empirically with surface temperature as a proxy for spatially dependent atmospheric processes.

However, the assumption that precipitation is dominantly controlled by evaporation via sunlight striking water cannot accurately represent carbon cycle feedback on arid planets; sunlight-driven evaporation would vastly underestimate the precipitation rate since evaporation would sharply decrease as the surface water reservoir drops. But even if small surface water reservoirs are not directly absorbing much installation, there is still energy in the system that can evaporate water, such as sunlight hitting land and driving planetary winds. Conversely, using the energy from all sunlight hitting the planetary surface will overestimate evaporation since only a fraction of this energy will be transferred to small bodies of surface liquid water.

For the exploration of arid terrestrial planets, we add wind-driven evaporation as a limit to precipitation (R. T. Pierrehumbert 2010). To determine the amount of evaporation (=precipitation) in our model, we calculate the latent heat flux as a function of temperature and scale it by a surface liquid water fraction. First, the saturation vapor pressure is calculated using the Clausius–Clapeyron relation:

$$p_{sat}(T) = p_{sat}(T_o) \, e^{-\frac{L_v}{R_W}\left(\frac{1}{T} - \frac{1}{T_o}\right)} \quad (14)$$

where $p_{sat}(T)$ [Pa] is the calculated saturation vapor pressure, $p_{sat}(T_o)$ [Pa] is the saturation vapor pressure at a reference temperature ($T_o = 273.15$ K), $R_W$ [J kg$^{-1}$ K$^{-1}$] is the specific gas constant for the condensing gas (water vapor, 461.5 J kg$^{-1}$ K$^{-1}$), and $L_v$ [J kg$^{-1}$] is a constant latent heat of vaporization of water ($2.5 \cdot 10^6$ J kg$^{-1}$ at $T_o = 273.15$ K). With $p_{sat}$, we can calculate the characteristic latent heat flux $F_L^*$ [W m$^{-2}$] associated with the transfer of mass from the surface to the overlying atmosphere:

$$F_L^* \equiv C_D U p_{sat}(T). \quad (15)$$

Here, $C_D$ is the drag coefficient, and $U$ is the mean horizontal wind speed; we assume a typical Earth-like average wind speed of 10 m s$^{-1}$, and an Earth-like, dimensionless drag coefficient for turbulent wind (0.001). $p_{sat}$ is calculated at each surface temperature $T$ [K]. Finally, we calculate the evaporative heat flux $E_o$ [W m$^{-2}$] as

$$E_o = (1 - h_{sa})\frac{L_v}{R_w T} F_L^* \quad (16)$$

where $h_{sa}$ is the relative humidity, and we use modern Earth's global average of 0.7. Varying the drag coefficient, wind speed, and relative humidity by factors of a few does not significantly affect the qualitative outcomes of our results. For a thorough description of latent heat flux during turbulent exchange, refer to R. T. Pierrehumbert (2010). We add the wind-driven evaporative heat flux to the MAC framework to calculate precipitation, $p$ [m yr$^{-1}$] scaled to the land fraction ($\gamma$), or rather the surface liquid water fraction, $(1 - \gamma)$:

$$p = \frac{E_o(T)(1 - \gamma)c_{yr}}{L_v \rho}. \quad (17)$$

To convert the evaporative heat flux precipitation to [m yr$^{-1}$], we divide by $L_v$ and the density of water $\rho$ (1000 kg m$^{-3}$) and multiply by a time unit conversion factor $c_{yr} = \frac{3.154 \cdot 10^7 \text{ s}}{1 \text{ yr}}$.

Figure 2 uses Equation (17) to calculate the precipitation rate at four different temperatures [273, 300, 320, 370 K], which are represented by four lines with colors that scale with temperature. The precipitation rate scales as expected with a dwindling water supply: it approaches zero as the land fraction increases to 100% (or as ocean fraction decreases to 0%). This analytic parameterization is supported by comparisons to two GCM studies of arid planets with varying water inventories in Figure 2 (e.g., M. J. Way & A. D. Del Genio 2020; D. M. Glaser et al. 2025). GCM results of arid planets broadly agree with wind-driven evaporation flux as a limit to precipitation. Crucially, when surface water inventory is low, there is an upper limit to silicate weathering set by precipitation and runoff. There are discrepancies with temperature since the GCM runs sample a large range of insolations and rotation rates, which also affect precipitation. With the addition of wind-driven evaporation, we can now assess the effects of shallow oceans on weathering fluxes.

### 2.2.4. Ingassing Flux

To calculate the final MAC weathering flux with wind-driven evaporation, the results from Equation (17) are used to calculate runoff using $q = \Gamma \cdot p$, where $\Gamma$ is the fraction of precipitation that becomes runoff, which is assumed to be fixed at 0.2. The resulting runoff parameterization is inserted into Equation (11), and the updated weathering rate $w$ is scaled by the planetary surface area and land fraction ($\gamma$) to calculate the ingassing flux of $CO_2$:

$$F_{CO_2}^{ingas}\left[\frac{\text{kg}}{\text{s}}\right] = M_{CO_2} \cdot w \cdot \frac{A_p}{c_{yr}} \cdot \gamma. \quad (18)$$

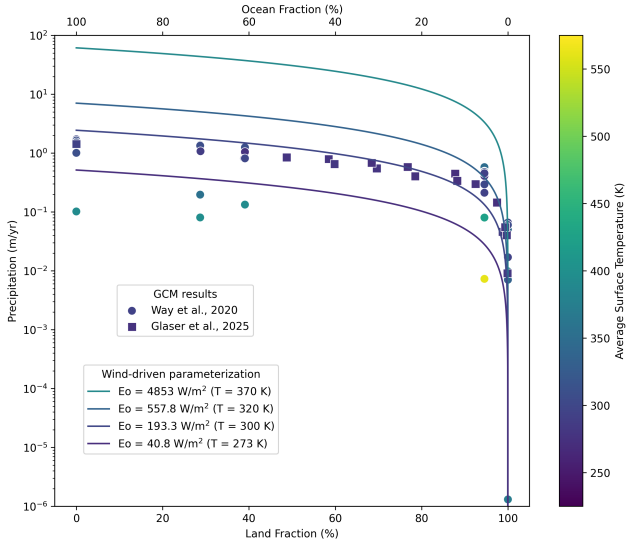

**Figure 2.** Precipitation estimated analytically for different surface temperatures assuming wind-driven evaporation limits (lines, R. T. Pierrehumbert 2010) is compared to the weighted average of precipitation from GCM runs of arid planets (dots, M. J. Way & A. D. Del Genio 2020; and square markers, D. M. Glaser et al. 2025). The average surface temperature associated with the calculated wind-driven precipitation and the GCM outputs are represented by a color heatmap. The M. J. Way & A. D. Del Genio (2020) GCM models tested five discrete land fractions, resulting in the five distinct dot groups. Broadly speaking the analytic precipitation-limited estimate of precipitation agrees with or overestimates precipitation from the GCMs.

$M_{CO_2}$ [kg mol$^{-1}$] is the molar mass of carbon dioxide, and $A_p$ [m$^2$] = $4\pi r^2$ is the planetary surface area ($r = 6.371 \cdot 10^6$ m for Earth). The weathering flux in our model now accounts for precipitation limits on arid planets. See Table 1 for all values assumed in our weathering model. Now that the weathering rate, $w$, is sensitive to temperature, $CO_2$, and runoff, we can use our updated $F_{CO_2}^{ingas}$ to calculate the surface and interior reservoirs of carbon, as described in Equations (3) and (4).

### 2.2.5. Land Fraction

The planet's land fraction is a function of the volume of surface water and planetary hypsometry. After the climate model determines the surface liquid water inventory, a hypsometric curve for Earth or Venus (R. Ghail 2015) is used to calculate how much of the planet would be submerged by that volume of water. Venus's topography is primarily shaped by the convecting mantle below, and any rocky planet with interior convection may exhibit this type of topography (C. M. Guimond et al. 2022). On Earth, this process is overshadowed by plate tectonics, resulting in a potentially rare hypsometry on the planet. Therefore, Venus's hypsometric curve might be a better representation for terrestrial exoplanets. We interpolate the hypsometric curve to create a smooth function that estimates elevation across finely spaced cumulative area bins. Starting from the lowest-elevation bins, we sum the water volume needed to flood each elevation slice until the total exceeds the available surface water. The point at which this occurs determines the fraction of the planet above sea level, i.e., the land fraction. Our model ignores water in subsurface reservoirs and aquifers. However, the existence of subsurface inventories could potentially remove available surface water from the carbon cycle, making an unbalanced carbon cycle more likely.

### 2.3. Outgassing

Outgassing ($F_{H_2O}^{volc}$ and $F_{CO_2}^{volc}$) is parameterized using the VolcGases model from N. Wogan et al. (2020). To summarize, their model describes the equilibrium exsolution of carbon, hydrogen, and oxygen-bearing volatiles from a magma source. The resulting fugacity of volatile species is governed by their solubility and the oxygen fugacity of the magma, assuming the system is in thermodynamic equilibrium. The model estimates the amount of $CO_2$ and $H_2O$ outgassed for a given magma temperature ($\sim$1700 K), pressure (calculated at each time step), oxygen fugacity ($\log fO_2 \approx $ QFM $- 1$), and mantle mass fraction of volatiles determined by $R_{H_2O}^{silic}$ and $R_{CO_2}^{silic}$. For a given melt production rate, VolcGases can be used to calculate volatile outgassing fluxes (e.g., J. Krissansen-Totton et al. 2021b, 2021c; J. Krissansen-Totton & J. J. Fortney 2022; A. O. Warren & E. S. Kite 2023).

Furthermore, we relate outgassing to internal heat flow and crustal production evolution relative to modern Earth. We consider both constant crustal melt production, and melt production declining exponentially, representative of a cooling planet (G. F. Davies 2009; M. Li et al. 2016) as two endmember cases. For the latter, internal heat flow relative to modern is given by

$$Q = \left(1 - \frac{t_0}{4.5 \cdot 10^9 \text{ yr}}\right)^{-n_{out}}. \quad (19)$$

$Q$ is the internal heat flow relative to modern Earth, and $n_{out}$ is an exponent that parameterizes Earth's heatflow evolution (see supplementary material for J. Krissansen-Totton et al. 2018). We consider a broad range of outgassing histories (Y. Godderis & J. Veizer 2000; N. H. Sleep & K. Zahnle 2001; J. Korenaga 2008; G. F. Davies 2009; see Table 2), by exploring both relatively constant heat flow through time (J. Korenaga 2008) and secular cooling (G. F. Davies 2009). Note that, within the forward model, $t_0$ [yr] moves backward in time to 4.5 Gyr and is given by $t_0 = 4.5 \cdot 10^9$ yr $- \frac{t}{c_{yr}}$.

We relate heat flow to the global crustal production rate (kg s$^{-1}$) by

$$F_{melt} = \rho_m \times V_{melt} \times Q^m \quad (20)$$

with the mantle density $\rho_m$ (4000 kg m$^{-3}$), modern melt volume rate $V_{melt}$ ($\sim$633 m$^3$ s$^{-1}$; M. Li et al. 2016), and $m$ is an exponent that determines the relationship between heatflow and crustal production (see Table 2). The VolcGases (functions $f$ and $g$) outgassing model (N. Wogan et al. 2020) then uses melt production, $F_{melt}$ [kg s$^{-1}$], to calculate the total outgassing flux at each time step in the model:

$$F_{H_2O}^{volc} = f(F_{melt}, R_{H_2O}^{silic}, R_{CO_2}^{silic}) \quad (21)$$

$$F_{CO_2}^{volc} = g(F_{melt}, R_{H_2O}^{silic}, R_{CO_2}^{silic}). \quad (22)$$

### 2.4. The Deep-water Cycle

The transfer of water into the interior ($F_{H_2O}^{ingas}$ in Equations (1) and (2)) is described by our formulation of the deep-water cycle. The deep-water cycle describes the exchange of water between the interior and the surface of a planet that is dictated by the competition between water ingassing and water outgassing fluxes (L. Rüpke et al. 2006).

**Table 2**
Uncertain Parameter Ranges Sampled in Carbon Cycle Calculations

| Parameter | Nominal Value | Monte Carlo Range | References/Notes |
|---|---|---|---|
| Initial surface water[a], $R^{\text{surf}}_{\text{H}_2\text{O}(t=0)}$ [kg] | $1.4 \times 10^{21}$ | $1.4 \times 10^{18}$–$1.4 \times 10^{21}$ | 0.1%–100% of Earth's oceans |
| Initial interior water[a], $R^{\text{silic}}_{\text{H}_2\text{O}(t=0)}$ [kg] | $0.5 \times 10^{20}$ | $10^{19}$–$10^{21}$ | J. Yang & M. Faccenda (2023) |
| Max interior storage capacity[a], $\Omega^{\text{silic}}_{\text{H}_2\text{O}}$ [kg] | $6.77 \times 10^{20}$ | $10^{20}$–$10^{22}$ | M. Hirschmann (2006); Z. Dong et al. (2019); C. M. Guimond et al. (2023) |
| Initial total carbon[a] $R^{\text{silic}}_{\text{CO}_2(t=0)} + R^{\text{surf}}_{\text{CO}_2(t=0)}$ [kg] | $10^{21}$ | $10^{20}$–$10^{22}$ | R. A. Fischer et al. (2020) |
| Carbon partitioning [%] $\frac{R^{\text{silic}}_{\text{CO}_2(t=0)}}{R^{\text{silic}}_{\text{CO}_2(t=0)} + R^{\text{surf}}_{\text{CO}_2(t=0)}}$ | 99[b] | 1–80 | Describes how much initial carbon is partitioned into the interior |
| Soil age[a] $t_s$ [yr] | $10^4$ | $10^3$–$10^5$ | K. Maher & C. P. Chamberlain (2014) |
| Kinetic weathering temp. dependence $T_e$ [K] | 11.1 | 5–15 | R. A. Berner (1994) |
| Internal heat flow exponent $n_{\text{out}}$ | 0.4 | 0–0.73 | G. F. Davies (2009); J. Korenaga (2008) |
| Outgassing exponent $m$ | 1.5 | 1.0–2.0 | J. Krissansen-Totton et al. (2021a) |

**Notes.** Columns denote parameter names (column (1)), nominal values used in single model runs (column (2)), the range sampled in Monte Carlo calculations (column (3)), and references or explanation for assumed value(s) (column (4)). Unless stated otherwise, parameters are sampled uniformly in linear space.
[a] Uniformly sampled in base-10 logarithmic space.
[b] Nominal value of 99% chosen to illustrate carbon cycle imbalance under the most optimistic assumptions (virtually all carbon starting in mantle). Our nominal range of 1%–80% represents more realistic starting conditions where most $CO_2$ is degassed during the magma ocean phase. Sampling 1%–99% partitioning in Monte Carlo runs does not substantially change results (not shown).

Thus, it is highly dependent on a planet's tectonic history (J. Korenaga et al. 2017; K. S. Karlsen et al. 2019). Although critical to regulating global climate and continental crust production, several aspects of this cycle are poorly understood. Previous studies have attempted to constrain the water storage capacity of the mantle (M. Hirschmann 2006; J. Dong et al. 2021; C. M. Guimond et al. 2023) and the upper limit of slab dehydration (R. Parai & S. Mukhopadhyay 2012; T. Garth & A. Rietbrock 2014; J. Yang & M. Faccenda 2023; G. S. Epstein et al. 2024). Estimates of the maximum mantle water capacity vary widely from 0.5 times the mass of Earth's modern oceans (K. Chotalia et al. 2023), to 15 times the amount of water on Earth's surface (B. Marty 2012). Deep-water cycling is even more uncertain for other solar system terrestrials and exoplanets; for example, constraining the extent of crustal hydration on Mars and the history of water loss (E. Chassefière et al. 2013; E. L. Scheller et al. 2021; D. Adams et al. 2025).

To account for these uncertainties, we broadly sample the maximum interior water storage (see Table 2 below) and adopt three end-member parameterizations to capture plausible deep-water cycle feedback: a "surface-dependent" case, a "mass-dependent" case, and a zero ingassing case. While the uncertainties in the deep-water cycle are not our focus, it is necessary to explore a plausible range of feedback because the deep-water cycle could potentially interact with carbon cycle feedback and surface habitability. The rate at which water is removed from the surface controls the water available for silicate weathering. If less water is on the surface due to a high ingassing flux, the carbon cycle may become unbalanced, even with a large initial surface water inventory. By exploring different plausible end-member cases for deep-water cycling, we will investigate the extent to which runaway $CO_2$ accumulation is an inevitable outcome on arid worlds.

In the surface-dependent case, the flux of water into the interior is calculated using a surface liquid water fraction:

$$F^{\text{ingas}}_{\text{H}_2\text{O}}(\text{surface-dependent}) = F_{\text{Earth}}$$
$$\times \frac{1 - \gamma}{1 - \gamma_{\text{Earth}}} \times \frac{F_{\text{melt}}}{F_{\text{modern}}} \times \max\left(0, 1 - \frac{R^{\text{silic}}_{\text{H}_2\text{O}}}{\Omega^{\text{silic}}_{\text{H}_2\text{O}}}\right) \quad (23)$$

where $F_{\text{Earth}}$ [kg s$^{-1}$] is the modern water flux to the interior, calculated as $8.9 \cdot 10^3$ kg s$^{-1}$, converted from the upper limit of water transport into the midmantle via deep slab subduction reported in J. Yang & M. Faccenda (2023) ($0.28 \cdot 10^{12}$ kg yr$^{-1}$). $1 - \gamma$ is the ocean fraction, and $1 - \gamma_{\text{Earth}}$ is Earth's modern ocean fraction (0.7), $F_{\text{modern}}$ [kg s$^{-1}$] is the modern melt production rate, $2.54 \cdot 10^6$ kg s$^{-1}$ (from M. Li et al. 2016, 20 km$^3$ yr$^{-1}$), and $\Omega^{\text{silic}}_{\text{H}_2\text{O}}$ is the maximum water storage capacity of the interior. For the surface-dependent end-member case, the rate at which water is removed from the surface depends on the surface area of water–rock contact, scaled by the melt production rate. This regime represents crustal hydration reactions as a function of the water–rock interface and is based on a similar parameterization in J. Krissansen-Totton et al. (2021b). The surface area of water available for water cycling is more sensitive to planetary hypsometry; for example, the presence of deep ocean trenches on Earth.

In the mass-dependent case, the flux of water into the interior is instead based on the mass of liquid water at the surface:

$$F^{\text{ingas}}_{\text{H}_2\text{O}}(\text{mass-dependent}) = F_{\text{Earth}}$$
$$\times (R^{\text{surf}}_{\text{H}_2\text{O}} \cdot \text{liq}_{\text{frac}}(T)/m_{\text{Earth}}) \times \frac{F_{\text{melt}}}{F_{\text{modern}}} \times \max\left(0, 1 - \frac{R^{\text{silic}}_{\text{H}_2\text{O}}}{\Omega^{\text{silic}}_{\text{H}_2\text{O}}}\right) \quad (24)$$

where $R_{H_2O}^{surf} \cdot \text{liq}_{frac}(T)$ is the liquid portion of the surface water reservoir given by the climate model. $m_{Earth}$ is the modern surface water mass ($1.4 \times 10^{21}$ kg). In this case, the volume of surface water controls the interior water flux. This mechanism for water exchange was first described by J. F. Kasting & N. G. Holm (1992) to explain why Earth's modern ocean is the optimal volume (or depth) for maximum hydrothermal activity at midocean ridges, thereby resulting in a significant water flux to the mantle. Their proposed mechanism is based on a self-stabilizing, pressure-dependent feedback that controls the surface water mass, which is broadly equivalent to the mass-dependence we describe here. The ocean depth and strength of these hydrothermal systems adjust over time to balance Earth's water ingassing and outgassing fluxes. Past water cycle models used a similar water mass-dependent formulation to assess the habitability of super-Earth exoplanets (N. B. Cowan & D. S. Abbot 2014; L. Schaefer & D. Sasselov 2015).

Ingassing fluxes are additionally scaled by the water content of the interior relative to maximum interior water content, as shown in Equations (23) and (24). Modifying the ingassing fluxes by this multiplicative factor ensures that ingassing diminishes as it approaches the maximum water storage capacity of the interior. The maximum interior water content ($\Omega_{H_2O}^{silic}$) is an unknown Monte Carlo parameter (see Table 2 below). Nominal calculations explore a plausible $\Omega_{H_2O}^{silic}$ range for mantle water storage, but we additionally include sensitivity tests where $\Omega_{H_2O}^{silic}$ represents only crustal water storage.

Each ingassing parameterization represents different deep-water cycle regimes, and thus differing outcomes for $R_{H_2O}^{silic}$ and $R_{H_2O}^{surf}$. These end-members bound plausible dependencies—if the water cycle is mass dependent, the ingassing flux will shrink dramatically for small water reservoirs, whereas the surface-dependent case preserves a high ingassing flux for shallow oceans.

We also consider a third end-member with no water ingassing and an intermediate case where ingassing is limited by the maximum water storage of the crust (sensitivity test described in the supplementary materials, Equations (A1)–(A6)). Limiting the depth of hydration to the crust prohibits water storage in the mantle, which may be the case on the Earth (J. Wade et al. 2017). The addition of these ingassing scenarios allows us to assess the importance of the deep-water cycle on the balance of the geologic carbon cycle.

### 2.5. Initial Conditions and Monte Carlo Simulations

We used Monte Carlo simulations to explore the sensitivity of our results to uncertainties in the geologic carbon cycle, deep-water cycle, and initial conditions. Several uncertain parameters and initial conditions were tested over a broad parameter space (Table 2). We ran the model 10,000 times and sampled all unknown parameters. The uncertain parameters sampled log-uniformly include initial surface water, initial interior water, initial total carbon, maximum interior water capacity, and soil age $t_s$. Uniformly sampled parameters include initial atmosphere–interior carbon partitioning, the temperature dependence of kinetic weathering $T_e$, the internal heat flow exponent $n_{out}$, and the outgassing exponent $m$.

### 3. Results

We first explore the effect of initial surface water inventory on the carbon cycle with individual model runs representing an Earth-analog (Earth-like insolation, hypsometry, and albedo), and nominal model parameters (Table 2). Figure 3 shows the time evolution from our coupled carbon cycle and deep hydrological cycle model. These outputs assume a mass-dependent deep-water cycle, constant melt production, and include diffusion-limited escape. Nominal parameter values are assumed, and results are shown for four initial surface water inventories (0.1%, 1%, 10%, and 100% of Earth's oceans). The subfigures show the time-evolution of surface water (Figure 3(A)), surface temperature (Figure 3(B)), precipitation (Figure 3(C)), atmospheric $CO_2$ (Figure 3(D)), carbon dioxide fluxes (Figures 3(E) and (G)), and water fluxes (Figures 3(F) and (H)).

For Earth-like initial water cases (10%, 100% of Earth's oceans; darker shades of blue in Figure 3), the carbon cycle remains balanced, and surface temperatures are temperate on geologic timescales as the silicate weathering feedback buffers against secular stellar evolution (Figure 3(B)). Precipitation slightly increases with time as the surface warms (Figure 3(C)), and $pCO_2$ decreases over time due to the carbonate–silicate weathering thermostat (Figure 3(D)). Outgassing and ingassing of $CO_2$ are also relatively stable (Figure 3(E)). The surface water inventory is approximately constant over the 5 billion yr evolution (Figure 3(A)), and so, weathering is never precipitation limited. Outgassing of $H_2O$ slightly increases, and $H_2O$ ingassing decreases with time (Figure 3(F)). This atmospheric evolution is comparable to what has been reported previously in carbon cycle models of Earth's long-term evolution (J. Krissansen-Totton et al. 2018).

Figure 3 also demonstrates how the carbon cycle evolution of arid planets diverges from that of Earth. Planets with low initial surface water inventories (0.1%, 1% of Earth's oceans, lighter shades of blue) enter a regime where silicate weathering cannot keep up with degassing, resulting in an unbalanced carbon cycle and triggering runaway warming. As stellar luminosity increases, the surface warms (Figure 3(B)), leading to enhanced evaporation and more water vapor in the atmosphere. With less liquid water remaining on the surface, precipitation rates decrease (Figure 3(C)). This occurs because crustal hydration gradually lowers surface water inventories (Figure 3(A)), and consequently, precipitation fluxes begin to drop because wind-driven evaporation is proportional to the surface ocean fraction (Figure 3(C)). At the same time, surface warming from increasing insolation partitions more surface water into the atmosphere, further limiting evaporation (=precipitation) fluxes. As precipitation falls, so too do silicate weathering fluxes, which become severely runoff limited (Figure 3(E)). This results in an unbalanced carbon cycle whereby outgassing exceeds weathering, causing $CO_2$ to accumulate in the atmosphere (Figure 3(D)), causing further surface warming (Figure 3(B)) and accelerating this process. Eventually, a critical threshold is crossed whereby all surface water is in the vapor phase, precipitation stops completely, and $CO_2$ and temperature runaway to high values. The increasing pressure overburden from $pCO_2$ suppresses replenishing $H_2O$ degassing (Figure 3(F)).

We repeated the calculations above with the alternative, surface-dependent water ingassing parameterization (supplementary, Figure A1). The qualitative behavior is the same, except that we observe a transition from a balanced to an imbalanced carbon cycle for larger initial water inventories somewhere between 10% and 100% of Earth's oceans. Low surface water inventory planets are vulnerable to additional water loss through atmospheric escape. Results that exclude escape processes are shown in a supplementary figure, Figure A2, and exhibit the same qualitative behavior as

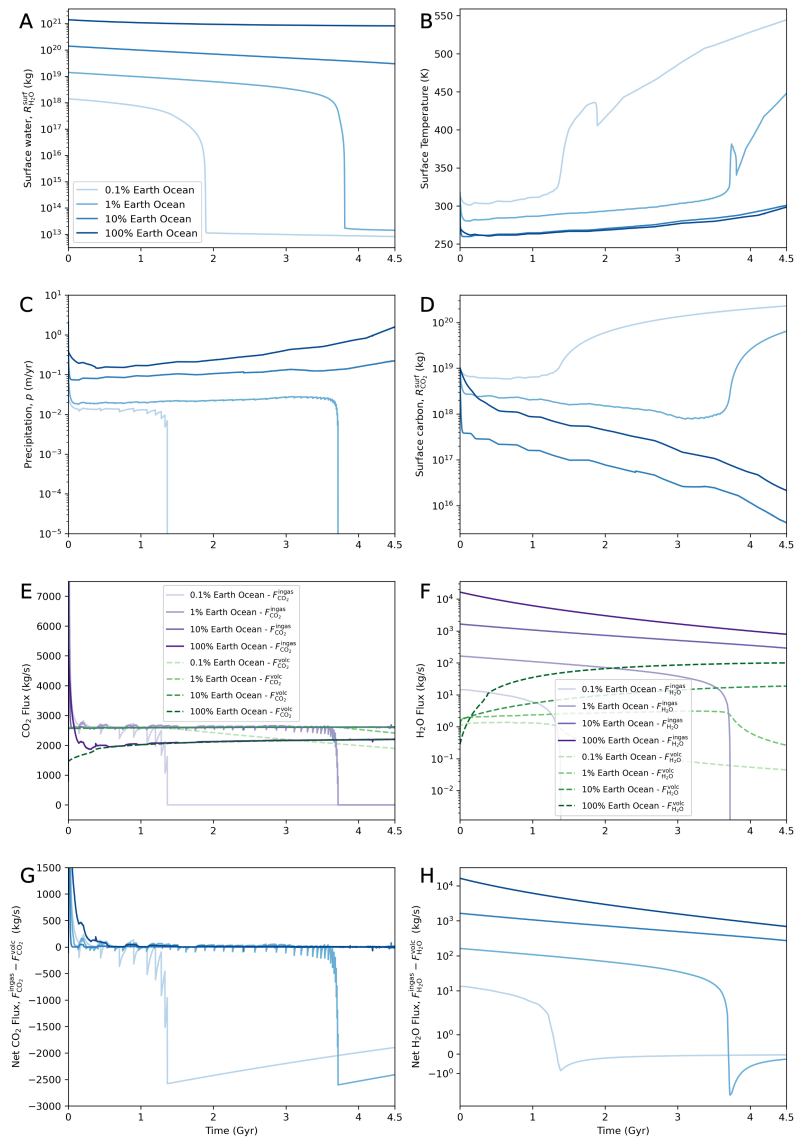

**Figure 3.** Four model outputs with varying initial water inventories denoted by blue shading. These outputs use a mass-dependent deep-water cycle, an albedo ($\alpha$) of 0.3, and escape processes are included. (A) Mass of all surface water reservoirs over the planet's 4.5 Gyr evolution. (B) Surface temperature. (C) Precipitation calculated from wind-driven evaporation. (D) Mass of all surface $CO_2$ including atmospheric $CO_2$ and dissolved carbon. The ingassing fluxes of $CO_2$ (E) and $H_2O$ (F) for various initial water inventories are denoted by purple shading. Outgassing fluxes of $CO_2$ (E) and $H_2O$ (F) are denoted by dotted green shading. (G) Net $CO_2$ flux into the interior. (H) Net $H_2O$ flux into the interior. In model runs with larger (10%, 100% of Earth's oceans) water inventories, surface temperatures are temperate through geologic time, but model runs with low initial water show runaway warming. Smaller (0.1%, 0.01% of Earth's oceans) initial water inventories have an unbalanced carbon cycle, increasing the flux of surface water into the interior while suppressing $H_2O$ outgassing. With less surface water, precipitation rates decrease (C), and silicate weathering fluxes become runoff limited, allowing volcanic outgassing to dominate (E). This results in surface warming (B) and increased concentrations of atmospheric $CO_2$ (D).

Figure 3. Without escape, planets can maintain a balanced carbon cycle at even lower water inventories, down to 1% of an Earth ocean on the surface (Figure A2).

### 3.1. Comparison of MAC versus WHAK Formulations

All model runs presented thus far use the MAC framework with wind-driven evaporation as a limit to precipitation described in Section 2.2. To illustrate that these carbon cycle instabilities are due to these new precipitation limits, we compare the MAC and WHAK formulations in Figure 4 for a planet with Earth-like instellation and hypsometry, mass-dependent water cycle and atmospheric escape. Without runoff sensitivity, the WHAK formulation produces habitable temperatures regardless of the initial water inventory. Even for an Earth-like planet with 0.1% of Earth's oceans, the final surface temperature is around 270 K. The MAC framework with wind-driven evaporation is sensitive to the amount of runoff and precipitation occurring on a planet, and shows a threshold of water required to maintain habitable surface temperatures between 1% and 10% of Earth's ocean mass.

### 3.2. Monte Carlo Sensitivity Calculations

Figure 5 plots the final surface temperature of 10,000 Monte Carlo runs as a function of their initial surface water mass for an Earth-like planet with $\alpha = 0.3$. The calculation was repeated for both a mass-dependent deep-water cycle and a surface-area-dependent deep-water cycle. For initial surface water inventories below about 20% of an Earth ocean, final surface temperatures were typically too hot to be habitable. We found that the mass-dependent case requires around 20% of Earth's ocean mass, but the surface-dependent case requires around 50% of an Earth ocean to reliably ensure temperate surface temperatures (probability $\approx 0$) after 4.5 Gyr of evolution. The surface-dependent case also shows a larger spread in final temperature values compared to the mass-dependent case. In both cases, there is a threshold of water needed to maintain a temperate climate.

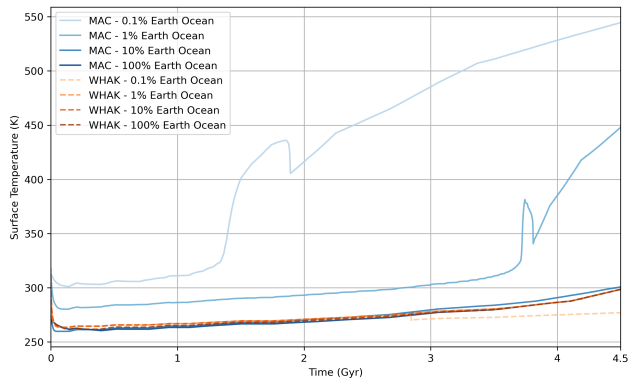

**Figure 4.** Comparison of WHAK and MAC weathering formulations with our geologic carbon cycle model for a planet with Earth-like hypsometry and instellation (Bond albedo, $\alpha = 0.3$). Mass-dependent water ingassing is assumed, and escape processes are included in these calculations. The WHAK model is denoted by shades of orange for varying initial water inventories, and the MAC model is denoted by shades of blue. Under the WHAK weathering formulation, the system maintains habitable temperatures even for very small initial surface water reservoirs. In contrast, the MAC model demonstrates a runoff dependency with runaway warming at low water inventories.

From the outputs in Figure 5, we calculate the fraction of model runs that end with final surface temperatures exceeding 400 K (i.e., probability $(T > 400 K)$), across bins of initial surface water mass. A threshold of 400 K is used to define the upper limit of habitability, as protein denaturation becomes significant at these temperatures (A. Clarke 2014). We use logarithmic binning to divide initial surface water values into 10 evenly spaced bins in log-space between 0.1% of Earth's oceans ($1.4 \cdot 10^{18}$ kg) to 100% of Earth's oceans ($1.4 \cdot 10^{21}$ kg). The probability that a model run ends with nonhabitable conditions is shown in Figure 6. In these plots, lower probabilities correspond to more habitable conditions, while higher probabilities indicate a greater likelihood that the model ends with surface temperatures exceeding 400 K. For example, a single run endowed with 10% of an Earth ocean (mass-dependent case) has a 20% chance of exceeding 400 K. As illustrated in Figure 5, the mass-dependent case achieves habitable conditions with less water, around 15% of Earth's ocean mass, whereas the surface-dependent case requires 30% of an Earth ocean to reduce the probability to 10% ($P = 0.1$).

We also tested scenarios with no ingassing (Figure 6), and where the surface- and mass-dependent end-member cases can only hydrate the crust (not the mantle) in Figures A4 and A5. To limit ingassing to the crust, we set the maximum interior capacity in our Monte Carlo model to 0.1%–10% of Earth's oceans ($10^{18}$–$10^{20}$ kg), calculated from the dehydration depth limit set by the geothermal gradient and lithostatic pressure (see supplementary materials, Equations (A1)–(A6)). Even with restricted or zero ingassing, the probabilities of runaway $CO_2$ accumulation are similar to the mass-independent ingassing case, indicating that arid planet carbon cycle feedback is largely independent of deep-water cycle assumptions.

We also explored how the other uncertain Monte Carlo parameters affected the probability of habitable surface temperatures in a supplemental figure, Figure A6, and found a positive relationship between the final surface temperature and the initial carbon reservoir, both on the surface and in interior. If the planet is endowed with more carbon, whether in the interior or on the surface, higher $CO_2$ leads to greater chances of higher temperatures. Larger $CO_2$ inventories lead to larger $CO_2$ degassing fluxes, resulting in a higher steady state of $pCO_2$ at any given weathering flux. Larger $CO_2$ degassing fluxes also necessitate higher precipitation and runoff rates to maintain a balanced carbon cycle, making it even more difficult for arid planets to sustain habitable conditions. There is also a negative relationship between the total water inventory and surface temperatures, but this is largely controlled by the initial surface water. For the other Monte Carlo variables (initial mantle water, soil age, kinetic weathering dependence, maximum interior water capacity, and outgassing parameters), there is no clear correlation with final temperature or the long-term climate balance. Instead, the initial surface water inventory and the precipitation limit to silicate weathering are the dominant controls for surface temperature evolution.

### 3.3. Venus-like Planets

We also investigated the balance of the geologic carbon cycle on planets with Venus-like hypsometry, instellation, radius, and mass. The effects of different albedo conditions are shown in Figure 7; for comparison, we also show an Earth-like planet with higher albedo. We use a Bond albedo of 0.65 at Venus orbit, which is approximately equivalent to an albedo of 0.3 for an Earth orbit. This represents the optimistic planetary albedo in M. J. Way & A. D. Del Genio (2020), which reports values up to 0.68 resulting from the water cloud feedback on the dayside of slow-rotating planets. Additionally, we include an Earth-like planet with an albedo of 0.65. Unsurprisingly, higher albedo conditions require less surface water to maintain habitable surface temperatures compared to similar models with lower albedo. Earth-like planets with $\alpha = 0.65$ can maintain temperate surface temperatures with as little as 5% of Earth's ocean mass of the initial surface water. For the high Venus albedo case presented in M. J. Way & A. D. Del Genio (2020), $\alpha = 0.65$, the probability of habitability is comparable to an Earth-like planet with $\alpha = 0.3$. Deep-water cycle assumptions—while still not affecting qualitative outcomes—do have a modest impact on the habitability threshold for Venus-like cases. At $\alpha = 0.65$, the mass-dependent case (dotted and solid light orange lines) has a higher probability of habitable surface temperatures than the surface-dependent case.

Furthermore, we plotted the time distribution of when the Venus-like model runs first exceeded the 400 K threshold in a supplementary figure, Figure A7, to measure when they transition to uninhabitable. Venus-like planets with mass-dependent ingassing enter a runaway warming state after an average of 0.577 Gyr while those with surface-dependent ingassing reach runaway slightly later, at an average of 1.07 Gyr. Thus, even under optimistic assumptions about albedo feedback, Venus-like arid planets often rapidly transition to uninhabitable. Only the surface-dependent ingassing case has a long tail of longer runaway transition times.

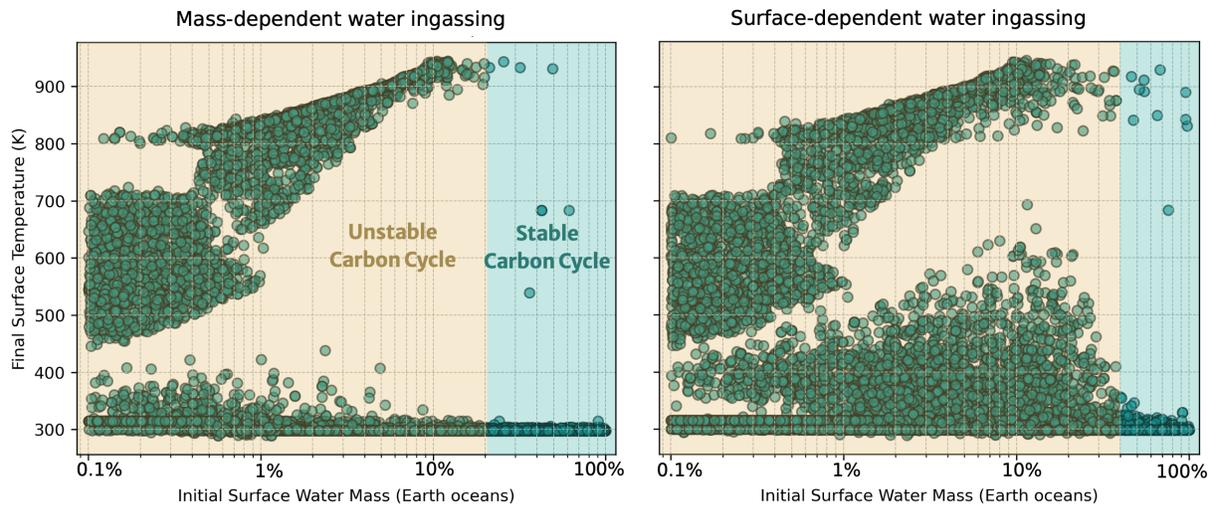

**Figure 5.** Planetary surface temperatures after 4.5 Gyr of model evolution as a function of the initial water inventory for Earth-like planets with $\alpha = 0.3$. Each dot represents the final surface temperature from our carbon cycle evolution model where we have performed a Monte Carlo calculation that randomly samples the initial surface and interior water mass, total carbon inventory, initial carbon partitioning between the surface and interior, soil age, temperature dependence of weathering, and several outgassing parameters. Parameter ranges are shown in Table 2. Yellow shading indicates the approximate region of initial water inventories where a greater percentage of model runs ends with uninhabitable surface temperatures resulting from an imbalanced carbon cycle. Cyan shading denotes the region where most model runs end with habitable surface temperatures and maintain a balanced carbon cycle. For planets of Earth-like mass, hypsometry, and instellations, those with a mass-dependent deep-water cycle require less water to maintain a balanced carbon cycle (20% of Earth's oceans) compared to a surface-dependent parameterization (~50% of Earth's oceans).

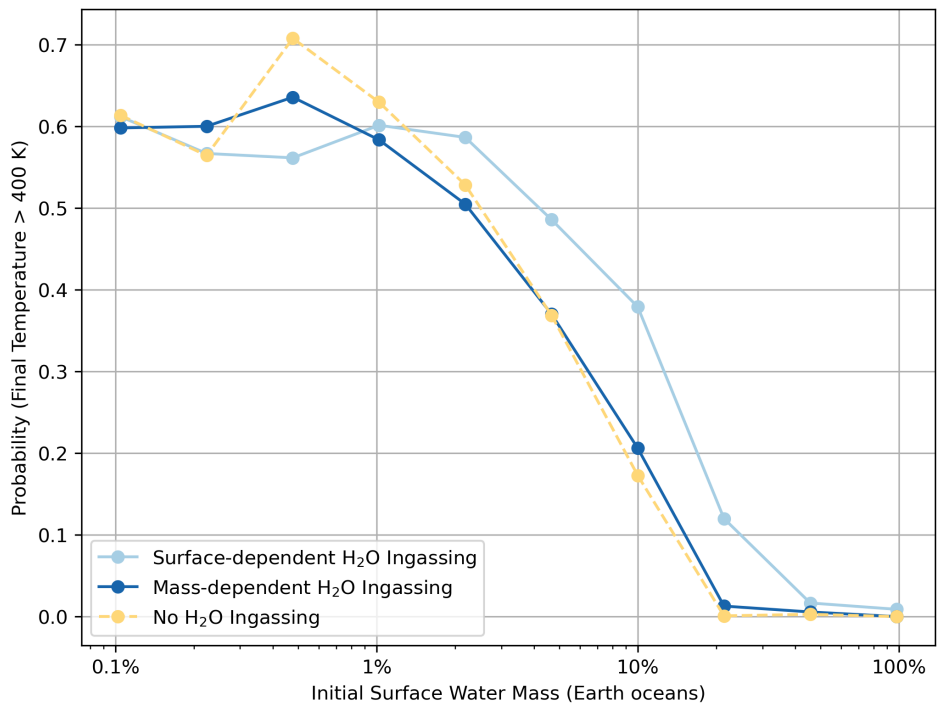

**Figure 6.** The probability of reaching a final surface temperature above 400 K for different initial surface water inventories at Earth-like instellation and hypsometry. Monte Carlo runs are divided into 10 evenly space logarithmic bins based on their initial surface water, ranging from 0.1%–100% of Earth ocean mass ($1.4 \cdot 10^{18}$–$1.4 \cdot 10^{21}$ kg). The bin center is calculated using the geometric mean of its edges. Lower probabilities correspond to more habitable conditions, as they are less likely to exceed the habitable temperature threshold of 400 K where proteins begin to denature (A. Clarke 2014). Surface- and mass-dependent parameterizations of the deep-water cycle are shown for an Earth-like planet (instellation, hypsometry) with $\alpha = 0.3$ (blue lines). Additionally, we include a third end-member where water ingassing is zero ("no ingassing," yellow line) meaning no water is transferred from the surface to the interior. The mass-dependent case is more likely to be habitable at lower initial water inventories, with ~15% of an Earth ocean marking the transition between a balanced and imbalanced carbon cycle evolution.

Overall, if Venus formed with an initial surface water inventory equal to or less than ~20%–50% of Earth's ocean mass, then any temperate early climate would have been susceptible to runoff-limited weathering and runaway warming. Compared to Earth, Venus is more likely to have had an unbalanced carbon cycle. If Venus formed with a similar (or slightly smaller) surface inventory to the Earth and had cloud feedback that maintained high albedos against increasing solar luminosity (M. J. Way & A. D. Del Genio 2020), then carbon cycle instabilities could have triggered a transition to Venus's modern, uninhabitable state. If water ingassing is surface dependent (darker orange lines), the likelihood of an imbalanced carbon cycle is even greater relative to Earth since Venus's hypsometry promotes more rock–water interaction under shallow ocean conditions.

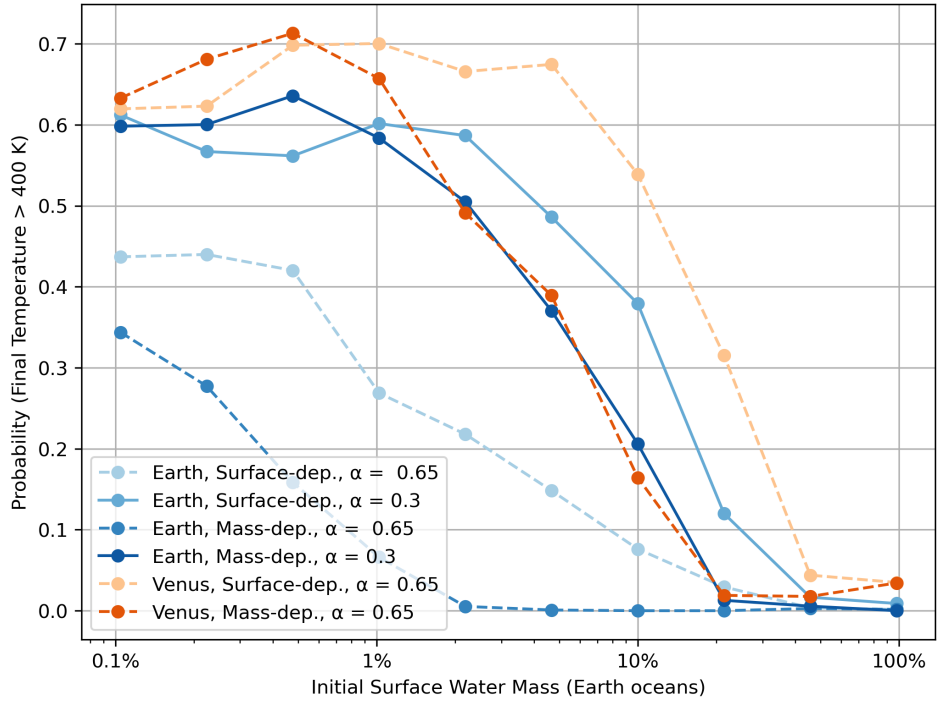

**Figure 7.** Probability of the Monte Carlo model runs reaching a final surface temperature of 400 K or higher. Similar to Figure 6, except showing the effect of albedo for planets with Earth-like and Venus-like hypsometry and instellation. Blue shades denote Earth-like planets, and orange shades denote Venus-like planets. Lighter shades of each color represent surface-dependent water ingassing, with darker shades representing mass-dependent water ingassing. Dashed lines represent models with $\alpha = 0.65$ for both Earth and Venus-like planets. Earth-like planets with higher albedo require less water to ensure balanced carbon cycles and retain habitable climates. However, Venus-like planets with high albedo and mass-dependent water ingassing show probabilities similar to Earth-like planets with modern albedo ($\alpha = 0.3$). Surface-dependent water cycles on Venus-like planets require more water to maintain temperate surface temperatures, potentially explaining the Earth–Venus dichotomy if Venus were ever habitable.

The Venus-like planets using the mass-dependent $H_2O$ ingassing (solid and dotted lines in darker shades of orange) exhibit their lowest probabilities of exceeding 400 K around 20% of Earth's ocean mass. However, the probability increases again as initial water inventories approach 100% of Earth's oceans. This rise is an unphysical artifact of excluding seafloor weathering in our model. At higher surface water inventories, it becomes easier to submerge Venus's topography due to the absence of a dichotomy in crustal topography. Venus's surface is relatively flat; 60% of its total surface is below 500 m of elevation (G. H. Pettengill et al. 1980), so even modest water depths can flood a large portion of the surface. To isolate the role of hypsometry, we also tested models with Venus-like instellation, but Earth-like hypsometry, as shown in Figure A8. As expected, the mass-dependent cases with Venus-like instellation (black and dark orange lines in Figure A8) remain more similar across hypsometric changes, since this deep-water cycle parameterization is less sensitive to hypsometry.

However, their probabilities diverge at larger water inventories. With Venus-like instellation and hypsometry, the probability of exceeding 400 K increases as the land is inundated, whereas, with Venus-like instellation and Earth-like hypsometry, the probability remains low since it possesses a crustal dichotomy.

*3.3.1. The Effect of Atmospheric Escape*

In our model, the surface water reservoir on arid terrestrial planets is partially controlled by the atmospheric escape of $H_2O$, which is calculated using a diffusion-limited escape flux. We include atmospheric escape as it may be significant to the long-term balance of the geologic carbon cycle. To estimate the escape flux on Earth, we can use the equation for diffusion-limited escape of H from J. F. Kasting et al. (1993).

$$\Phi_{esc} = 2 \cdot 10^{13} \cdot 2f(H_2O) \quad (25)$$

where $f(H_2O)$ is the total water mixing ratio in the stratosphere. At 300 K, the escape rate is $6.34 \cdot 10^8 \, cm^{-2} \, s^{-1}$. Applying this escape rate over the surface area of the Earth results in 0.44% of Earth's oceans of water escaping over 4 Gyr, which overlaps with the range of initial water inventories tested here. As temperatures rise during an unbalanced carbon cycle, the escape rate at 350 K results in the loss of nine Earth's oceans over 4 Gyr. A significant amount of surface water is lost over a planetary timescale from diffusion-limited escape, and this water loss accelerates as temperatures rise.

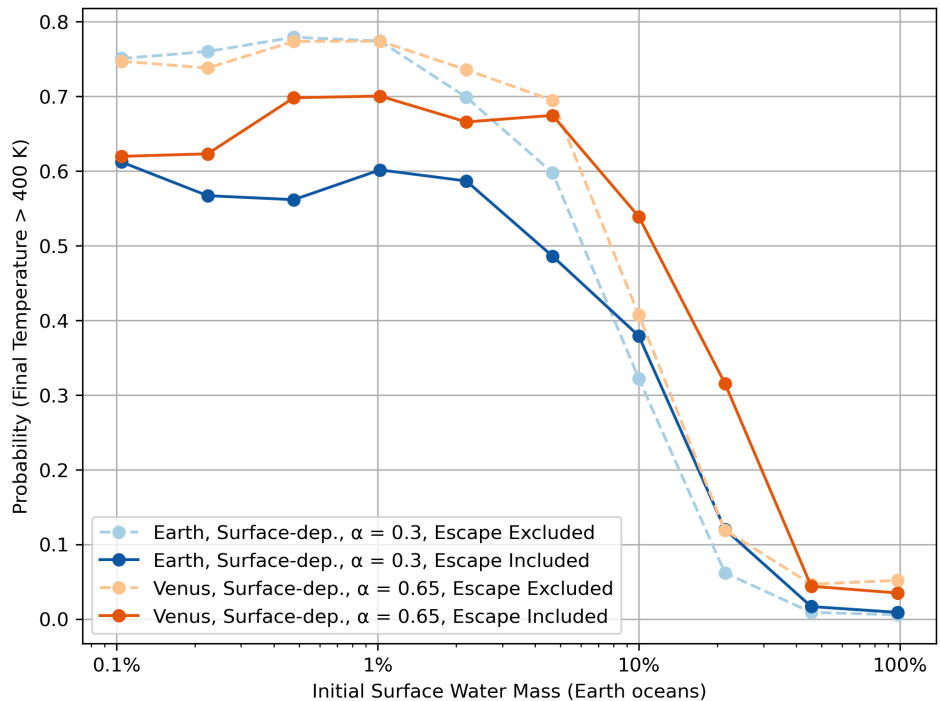

**Figure 8.** Probability plot showing the effect of escape processes at Earth-like and Venus-like instellations. Venus-like hypsometry and instellation is denoted by orange shades, and Earth-like models are in shades of blue. All models use surface-dependent water ingassing cycles. Runs that exclude atmospheric escape are denoted by the dotted lines and lighter shades. Atmospheric escape is less significant to habitability outcomes at higher initial water inventories above 10% of Earth's ocean mass.

However, the rate of water loss to space is limited by the atmospheric cold trap (D. C. Catling & J. F. Kasting 2017). If temperatures are too cool, then rising water vapor will condense, decreasing $H_2O$ transport to the upper atmosphere, where $H_2O$ photolysis and subsequent H escape can occur. R. D. Wordsworth & R. T. Pierrehumbert (2013b) found that $CO_2$ can enhance the atmospheric cold trap through efficient radiative cooling in the upper atmosphere, creating a bottleneck for planetary water loss. In our models, balanced carbon cycle regimes show decreasing surface $CO_2$ concentrations over time (Figure 3(D), darker blue shades), but imbalanced carbon cycle climates result in $CO_2$ accumulation (Figure 3(D), lighter blue shades), which may limit the extent of H escape. Therefore, our modeled atmospheric escape processes may overestimate escape on arid planets with unbalanced carbon cycles. To address this, we modeled the geologic carbon cycle with and without atmospheric escape processes.

All model outputs presented above include atmospheric escape. Here, we repeat selected simulations without atmospheric escape to test the extent to which water loss is driving our results. Overall, we find that escape has a small impact on our conclusions. Figure 8 compares the probability that the model runs exceed 400 K for both Venus and Earth-like planets with surface-dependent ingassing with and without escape. A supplementary figure, Figure A9, shows the results for models with mass-dependent ingassing with and without escape. Figure A5 shows the results for all of our ingassing parameterizations without escape (Figure A4 presents the ingassing parameterizations that include escape). At low initial water inventories, omitting escape actually increases the likelihood of the planet evolving toward uninhabitable surface temperatures. Model runs that exclude escape accumulate higher levels of water vapor and exhibit greater greenhouse warming compared to model runs that include escape processes. When initial water inventories exceed 10% of Earth's ocean mass, the impact of atmospheric escape becomes much less significant. At these higher initial inventories, the models without atmospheric escape require slightly less water to maintain habitable conditions than their counterparts that include escape. This is because atmospheric escape can accelerate the destabilization of the carbon cycle by enhancing surface water loss.

## 4. Discussion

Our results show that arid terrestrial planets may have unbalanced geologic carbon cycles due to runoff-limits to weathering, which can lead to a loss of habitability and runaway warming. Even if a planet is in the HZ, it can transition to an uninhabitable state if it does not have enough initial surface water to balance outgassing and weathering fluxes. With less available liquid water, $CO_2$ drawdown from continental weathering becomes limited by precipitation, allowing $CO_2$ to accumulate in the atmosphere and further warm the surface. Across our simulations, we find the minimum threshold of surface water required to maintain a balanced carbon cycle is around 0.2–0.5 Earth's oceans by mass. These results are robust against different values of initial interior water, maximum interior water capacity, total planetary carbon, weathering parameters, and planetary hypsometry. Therefore, arid planets with much less than one Earth ocean of surface water are unlikely to remain habitable. Previous work has shown that the topography and continental orientation on dry planets may provide a stabilizing climate feedback against the runaway greenhouse limit (Y. Abe et al.

2011; D. M. Glaser et al. 2025), but while this feedback may operate efficiently on short timescales, they might be insufficient over geologic time without a working carbonate–silicate weathering thermostat.

### 4.1. Deep-water Cycle Dependencies

We used three parameterizations for the deep-water cycle to assess its influence on the silicate weathering thermostat's ability to modulate climate (Figure 6). A variety of parameterizations of the deep-water cycle have been adopted to explore the volatile evolution of terrestrial exoplanets (J. F. Kasting & N. G. Holm 1992; N. B. Cowan & D. S. Abbot 2014; L. Schaefer & D. Sasselov 2015; T. D. Komacek & D. S. Abbot 2016; C. M. Guimond et al. 2023). Here, we tested the effects of three extreme end-member cases: a surface-dependent case that depends on water–rock contact, a mass-dependent case that depends on ocean volume, and a no ingassing of water scenario. In supplementary figures, Figures A4 and A5, we also consider an intermediate case where hydration of the interior is limited to the crust. The results of the "no ingassing" and "crustal hydration only" cases are similar to our mass-dependent ingassing formulation. Overall, these limited ingassing systems behaves similarly to both of our end-member cases. This is also demonstrated in Figure A6, where the initial interior water and maximum interior water capacity do not show a strong influence on the final surface temperature. Therefore, compared to the initial surface water inventory, the deep-water cycle is not a primary determinant of surface habitability. When the carbon cycle is imbalanced and the climate is in a runaway state, water loss to the interior or to space is even less significant for climate evolution.

Between our three end-member ingassing scenarios, mass-dependent water ingassing generally requires less water to maintain temperate conditions compared to the surface-dependent end-member. This arises because the mass-dependent formulation assumes that the flux of water into the interior scales with the total volume of surface water available, rather than the surface area of water–rock contact. In this regime, diminishing surface water leads to proportionally lower interior water return fluxes, which helps to preserve surface water inventories and maintain a balanced carbon cycle. In contrast, the surface-dependent formulation maintains high ingassing fluxes even at low surface water inventories, which accelerates surface desiccation and promotes an earlier onset of runoff-limits to weathering. As a result, across both the Earth-like and Venus-like models, the surface-dependent case required 2–4 times as much initial surface water to ensure habitable surface conditions. Model runs with higher albedo showed an even greater disparity in habitability outcomes between these end-member cases (Figure 7). Overall, the three end-member water ingassing scenarios resulted in some differences in their probabilities of habitable conditions, but the specific details do not change the overall conclusion that smaller surface water inventories are the dominant controls for long-term habitability.

Although our results did not find the deep-water cycle to be a first-order control for the evolution or arid terrestrial planets, substantial uncertainties remain concerning the exchange of water between the surface and the interior. On Venus and Earth, most of their bulk silicate water budgets may have been retained in the mantle shortly after magma ocean solidification, with upper limit estimates ranging from 77%–99% (D. J. Bower et al. 2022; Y. Miyazaki & J. Korenaga 2022; S. Hier-Majumder & M. M. Hirschmann 2017). Whether the transfer of water from the surface to the mantle was dictated by the surface area of available water or ocean volume has implications for whether rapid degassing from a hot mantle will offset water ingassing. On Mars, the details of crustal hydration are a key to reconstructing the size of ancient water inventories and how they were lost over time (E. L. Scheller et al. 2021; R. D. Moore et al. 2025). Although we find that very arid planets ($\ll 10\%$ of Earth's oceans) are unlikely to be habitable under any circumstances, the long-term habitability of planets with 10%–50% of Earth ocean mass may depend on the details of the deep-water cycle.

### 4.2. Implications for Rocky Exoplanets

Our results linking the size of the water reservoir with long-term climate will potentially be testable with the upcoming Habitable Worlds Observatory (HWO; National Academies of Sciences, Engineering, and Medicine 2023), which will constrain surface habitability and land fraction via reflected light spectroscopy. Detecting surface liquid water directly with HWO may be possible by measuring the specular reflection of starlight off an exoplanet's surface ocean, i.e., ocean glint (D. M. Williams & E. Gaidos 2008; D. J. Ryan & T. D. Robinson 2022; S. R. Vaughan et al. 2023) or measuring an exoplanet's rotational light curves and constructing a spatial map (N. B. Cowan et al. 2009). Combining both methods by mapping the ocean glint on a planet can provide even more confident ocean detections (J. Lustig-Yaeger et al. 2018). Furthermore, studies simulating reflected light retrievals from the HWO show that the future telescope may be able to measure the land fraction of exoplanets (A. G. Ulses et al. 2025). These retrievals would not rely on ocean glint but rather on the relative contributions of positively sloped (land) and negatively sloped (ocean/ice/snow) reflectance spectra.

The James Webb Space Telescope is currently observing terrestrial planets in M-dwarf systems that could have experienced aridity–carbon cycle feedback similar to that described here. M-dwarf systems are especially compelling due to their abundance and the observational favorability of close-in HZ planets. Terrestrial planets around M-dwarfs likely experience a slower magma ocean solidification phase due to their star's extended pre-main sequence (R. Luger & R. Barnes 2015; P. Barth et al. 2021; K. Moore et al. 2023; J. Krissansen-Totton et al. 2024). This results in significant surface and mantle desiccation due to thermal escape, but escape processes will slow as the water reservoirs decrease, especially if water migrates to the cooler nightside of the planet (F. Ding & R. D. Wordsworth 2020). Future work should explore the habitability prospects for these arid M-dwarf planets after the pre-main sequence.

### 4.3. Implications for Venus

Whether Venus ever possessed liquid water oceans and temperate climate conditions is a long-standing question in astrobiology and terrestrial planet evolution. With several Venus missions planned for the early 2030s (R. Ghail et al. 2021; J. B. Garvin et al. 2022; S. Smrekar et al. 2022), more evidence in favor of past habitability may be uncovered. Whether geologic investigations in Venusian tesserae

(M. S. Gilmore et al. 2015) or updated isotopic measurements from the atmosphere (G. Avice et al. 2022), any evidence for past clement conditions would demand an explanation for how habitability was lost. Likewise, evidence against past temperate conditions merits an explanation for why Venus never achieved long-term habitability, unlike Earth.

It has been hypothesized that a massive outgassing event, such as stochastic overlap of large igneous province eruptions or catastrophic global resurfacing, injected the atmosphere with $CO_2$ and triggered a transition to the modern runaway greenhouse conditions (M. J. Way & A. D. Del Genio 2020; M. J. Way et al. 2022). On water-rich planets, silicate weathering—operating on timescales of approximately $10^5$ to $10^6$ yr (E. T. Sundquist 1991; T. M. Lenton & C. Britton 2006)—can ostensibly stabilize the climate following substantial $CO_2$ injections, well before significant water loss to space occurs, which typically unfolds over timescales of $10^9$ yr for ocean-equivalent water masses (J. F. Kasting 1988). The silicate-carbonate weathering feedback can stabilize the climate against increases to outgassing so long as there is adequate supply of fresh, weatherable rock (B. P. Coy et al. 2025), but it cannot stabilize the climate against small water inventories and runoff-limited weathering (i.e., Figure 3(E)).

According to our models, the threshold of the initial surface water needed to sustain a balanced climate on Venus ranges from 20%–50%. The precise threshold of surface water needed to maintain temperate conditions depends on the water ingassing regime, albedo, and outgassing history. If Venus formed with a surface water inventory above this threshold, then it will probably have had enough precipitation to maintain a balanced geologic carbon cycle and habitable surface conditions for 4.5 Gyr (assuming cloud feedback to maintain a high albedo). However, if the initial water inventory fell below this range, then weathering would have become severely runoff limited and destabilized the carbon cycle, leading to runaway warming and higher concentrations of atmospheric $CO_2$. Even for the most optimistic high albedo case and a mass-dependent water cycle, planets at Venus-like instellations cannot maintain a balanced carbon cycle below 20% of Earth's oceans.

Determining the size of Venus's initial surface water inventory is one way to constrain its evolutionary history and loss of habitability. D/H measurements predict a past ocean mass 1.3%–17% of Earth's modern oceans (T. M. Donahue & C. T. Russell 1997), which would support a carbon cycle-driven climate transition, although early hydrodynamic H loss may have been nonfractionating, implying a much larger initial water inventory (K. J. Zahnle & J. F. Kasting 1986). Past surface water inventories have also been constrained by the absence of $O_2$ in the modern atmosphere, which is expected to accumulate following the photodissociation of $H_2O$ and the subsequent preferential escape of hydrogen to space. The extremely low oxygen abundance in the modern venusian atmosphere (V. I. Oyama et al. 1980) implies that, if oceans did exist, efficient oxygen sinks must have operated to remove the residual $O_2$. Estimating the efficiency of these oxygen sinks alongside rates of atmospheric escape is challenging, but estimates of Venus's maximum initial surface water inventory from this approach range from 10%–100% of an Earth ocean, but most estimates cluster around 20%–40% of an Earth ocean (A. O. Warren & E. S. Kite 2023; J. Krissansen-Totton et al. 2021a; C. Gillmann et al. 2009). These results may also be testable with isotopic measurements from future Venus missions. Neon isotopic ratios have been measured and provide some information on the initial volatile budget, although Kr and Xe measurements are needed to better constrain the initial water inventory and the early atmospheric loss processes on Venus (G. Avice et al. 2022; C. Gillmann et al. 2022). Ne ratios, in combination with Kr and Xe abundances and isotopes, may collectively constrain the delivery and loss of Venus volatiles. Altogether, these measurements could shed light on the available surface water inventory for carbon cycling and how water was lost over time.

### 4.4. Limitations of Climate Modeling

The primary limitations of our model involve the lack of spatial resolution and some of our assumptions regarding atmospheric structure. The results presented here are based on a 1D column climate model coupled with a 0D representation of the surface. This makes it difficult to assess the spatial distribution of the planetary system, in particular, the distribution of surface water and precipitation. In practice, the geographic location and extent of the surface water reservoir can dramatically affect the climate of terrestrial planets. Accounting for the spatial resolution further complicates the conditions needed for a balanced carbon cycle. 3D GCMs from F. Ding & R. D. Wordsworth (2021) found that synchronously rotating arid exoplanets tend to cold trap their limited surface water inventories on the nightside as ice caps, potentially limiting weathering fluxes even more than our globally averaged calculations suggest. Continental configuration on arid planets also plays a role in controlling surface temperatures and albedo from snow and ice deposition (D. M. Glaser et al. 2025). While our deep-water cycle model accounts for the ocean fraction, it does not resolve where precipitation occurs across a planetary surface. Consequently, even a planet with comparatively large surface water inventories may be unable to balance carbon fluxes if the water is confined to a few isolated basins and lakes where no new crust is forming. Future 3D modeling will be needed for the geologic carbon cycle to understand the importance of the spatial distribution of surface oceans.

Our climate model assumes an isothermal to moist adiabat to dry adiabat temperature profile (J. F. Kasting 1988) when computing surface temperature and liquid/gaseous surface water partitioning. This assumption works well for the relatively thin atmospheres considered here, but can break down for optically thick atmospheres with substantial $CO_2$ or steam. At high enough pressures, a full radiative–convective profile yields lower temperatures in comparison to the dry adiabatic profile (F. Selsis et al. 2023; J. Cmiel et al. 2025). However, for steam atmospheres, the differences in profile behavior are most significant above 1000 K or above ~50 bar; below this, the radiative–convective and dry adiabat profiles are comparable (F. Selsis et al. 2023). Our model's surface temperature evolution rarely reaches such high values. Therefore, the dry adiabat assumption is likely sufficient for our modeling approach, but future work could potentially explore carbon cycle dynamics with a full radiative–convective model, particularly during runaway $CO_2$ accumulation where atmospheric steam is increasing.

Since our current climate model does not track nitrogen fluxes, we assume a 1 bar partial pressure of $N_2$ ($pN_2$) for all model runs. Lower $pN_2$, as seemingly occurred on the early Earth (S. M. Som et al. 2016), would allow for higher rates of atmospheric escape, accelerating the loss of surface water. We have tested our model with and without escape (Section 3.3.1) to ensure that this arid carbon cycle feedback is primarily a result of surface water loss, not escape. Conversely, higher $N_2$ concentrations are possible on arid planets, such as on modern Venus with a $pN_2$ of ~3.2 bar (V. I. Oyama et al. 1980). Higher $N_2$ pressures could potentially warm the surface, accelerating the precipitation-limited feedback that ultimately results in a carbon cycle imbalance (A. Paradise et al. 2021). However, increasing $pN_2$ up to 3 times the modern Earth concentrations is not expected to have a modest impact on surface temperature (R. Wordsworth & R. Pierrehumbert 2013a; A. Paradise et al. 2021). Allowing for $N_2$ to vary is unlikely to dramatically change carbon cycle dynamics compared to our fixed 1 bar assumption, but this sensitivity to pressure could be explored in future work. The climatic and carbon cycle impact of $O_2$ not consumed by crustal sinks is similarly a topic for future exploration.

### 4.5. Future Opportunities for Spatially Resolved Modeling

Precipitation in our model is calculated from the latent heat flux on the surface and is scaled by land fraction. We use this precipitation in the MAC weathering formulation to ensure that precipitation decreases at higher land fractions, which is characteristic of these arid worlds. Since our model is 1D, it cannot account for spatial relationships modeled in 3D GCMs. This could potentially affect our conclusions if GCMs of arid planets predicted precipitation rates higher than our wind-driven evaporation parameterization. This would imply that precipitation is more efficient on arid planets than what we assume, potentially permitting a balanced geologic carbon cycle at low initial surface water inventories. However, compared to GCM outputs from M. J. Way & A. D. Del Genio (2020) and D. M. Glaser et al. (2025), our model typically overestimates the global precipitation rate (Figure 2). Furthermore, the expected concentration of precipitation in colder regions would imply slower weathering kinetics than globally averaged calculations would suggest, further reducing weathering fluxes (L. R. Kump & M. A. Arthur 1997; J. K. Caves et al. 2016). Spatially dependent precipitation estimates would exacerbate the problem of maintaining a balanced carbon cycle, and potentially, more surface water would be needed to maintain a balanced carbon cycle. We argue that the wind-driven parameterization described in Section 2.2.3 is conservative in that it may overestimate carbon cycle stabilization in arid climates.

With that said, modeling the 3D affects of precipitation will be critical in future work to further assess the habitability of these arid terrestrial planets. Our model assumes that precipitation scales directly with latent heat, which assumes that precipitation is tied to surface temperature and energy availability. But our 1D model does not capture the spatial decoupling between moisture and energy sources that can occur on 3D worlds. For example, Y. Abe et al. (2011) found that, on dry but otherwise Earth-like planets, most surface water is stable at the poles. On tidally locked arid planets, most surface water may be trapped on the cooler nightside of the planet since the atmosphere is optically thin to infrared and inefficient at transporting heat, resulting in an enhanced surface cold trap (F. Ding & R. D. Wordsworth 2021). Some arid M-dwarf terrestrials may only have habitable surface at the dayside–nightside terminator boundary (A. H. Lobo et al. 2023). In all such cases, precipitation may occur in regions with limited instellation or cooler temperatures, a behavior our model cannot capture.

Several studies have shown that the geographic distribution of precipitation and water sources is a key determinant of surface habitability (Y. Abe et al. 2011; T. Kodama et al. 2019; F. Ding & R. D. Wordsworth 2021; D. M. Glaser et al. 2025). In 3D planetary climates, atmospheric and oceanic circulation, topography, and moisture transport all contribute to global precipitation, not just temperature. The subsurface is also relevant to these hydrologic parameters, as subsurface water may transiently enhance weathering fluxes. On Earth, subsurface water can weather minerals deep in the crust, as shown by high bicarbonate concentrations measured in groundwater (S. Zhang & N. J. Planavsky 2019). Estimates of this subsurface silicate weathering flux indicate that groundwater weathering is potentially a significant carbon sink in long-term carbon cycling (S. Zhang & N. J. Planavsky 2019). On arid planets, however, the decreased precipitation fluxes would struggle to replenish the subsurface water reservoir, which would ultimately equilibrate with the surrounding rock. The work presented here does not model these 3D hydrologic processes, and instead provides a conservative estimate for the amount of the initial surface water needed to maintain the silicate-carbonate thermostat on terrestrial planets. Future work should study the geologic carbon cycle on arid planets in the 3D space to address these 1D model deficiencies.

### 5. Conclusion

We modeled the geologic carbon cycles on arid terrestrial planets to explore their prospects for habitability. Our results show that the carbon cycle's ability to maintain temperate surface conditions is largely controlled by the planet's initial surface water inventory. On Earth-like planets (instellation, hypsometry, albedo), around 20%–50% of Earth's ocean mass is needed to ensure habitable surface temperatures depending on the details of the deep-water cycle. If the planet is endowed with less than 20%–50% of an Earth ocean, then silicate weathering becomes severely runoff limited and unable to balance the outgassing flux of $CO_2$ into the atmosphere. In an unbalanced carbon cycle regime, $CO_2$ accumulates in the atmosphere, eventually leading to runaway warming and uninhabitable surface conditions. Although these planets are within their star's HZ, without enough surface water, they can still transition to an uninhabitable state.

In addition, we tested the carbon cycle's ability to maintain habitable temperatures at Venus-like instellations and optimistic albedos representing cloudy dayside negative feedback. Depending on the type of $H_2O$ ingassing, Venus-like planets require an initial water inventory of 20%–50% of Earth's oceans for the carbon cycle to remain balanced. An imbalanced carbon cycle climate transition can account for the current $CO_2$-dominated atmosphere and high surface temperatures (~450°C). If Venus were endowed with a similar or smaller surface water inventory than Earth, the resulting carbon cycle imbalance may have triggered a climate transition to its current inhospitable state around 1 Gyr after formation.

Our results show that arid terrestrial planets are unlikely to be habitable since the lack of surface water cannot maintain a balanced geologic carbon cycle over long timescales. Depending on the details of deep-water cycling, outgassing history, and albedo, a threshold of surface water is needed for surface habitability. This work may be testable with future exoplanet telescopes detecting land/ocean fraction and upcoming Venus missions measuring isotopic ratios that may constrain the initial surface water inventory and long-term atmospheric evolution.

## Acknowledgments

Thank you to Claire Guimond and another anonymous reviewer for the constructive feedback. We also thank Donald Glaser, Michael Way, and Robin Wordsworth for their helpful discussions and suggestions. H.W.-G. was supported by the National Science Foundation Graduate Research Fellowship Program (NSF-GRFP) under grant No. DGE-2140004. J. K.-T. was supported by the Virtual Planetary Laboratory, a member of the NASA Nexus for Exoplanet System Science (NExSS), funded via the NASA Astrobiology Program grant No. 80NSSC23K1398 and the Alfred P. Sloan Foundation under grant No. 2025-25204.

## Appendix
## Supplementary Results

### A.1. Nominal Results

Figure A1 shows the nominal model results for an Earth-like planet with surface-dependent ingassing. The overall behavior is similar to Figure 3 (mass-dependent ingassing), except that more surface water is required to prevent runaway warming, around 10%–100% of an Earth ocean.

Nominal model runs that neglected atmospheric escape for an Earth-like planet and mass-dependent water cycle are shown in Figure A2. Neglecting atmospheric escape helps the carbon cycle remain balanced, as only 0.1%–1% of an Earth ocean is needed to maintain clement temperatures. Atmospheric escape enhances water loss, which contributes to the destabilization of the carbon cycle.

### A.2. Limited Interior Hydration Sensitivity Test

To assess the importance of the deep-water cycle on the geologic carbon cycle, we tested the three end-member scenarios in the main text (surface-dependent, mass-dependent, and zero ingassing). Here, we consider the intermediate case where ingassing is restricted to crustal hydration only.

To limit interior hydration to the crust, we changed the maximum interior water storage ($\Omega_{H_2O}^{silic}$) from mantle values to crustal values to investigate a scenario where hydrated crust is not subducted into the mantle. We calculated the maximum depth of crustal hydration by finding the depth where the geothermal gradient intersects the dehydration temperature of serpentine. First, we calculated the surface heat flux ($q_{surface}$) throughout the planet's evolution:

$$q_{surface} [\text{W m}^{-2}] = q_{0,modern} \ Q^{n_{out}} \quad \text{(A1)}$$

where $q_{0,modern}$ is Earth's modern interior heat flux (0.08 W m$^{-2}$), $Q$ is the internal heat flow relative to modern Earth calculated from our outgassing model (see Section 2.3), and $n_{out}$ is constant (0.73). From the heatflow, we calculated the geothermal gradient ($\frac{dT}{dz}$) using Fourier's law of heat conduction:

$$\frac{dT}{dz} [\text{K m}^{-1}] = \frac{q_{surface}}{k_{th}}. \quad \text{(A2)}$$

$k_{th}$ is the thermal conductivity of the lower crust (2.5 W m$^{-1}$ K$^{-1}$; L. Ray et al. 2015). We then calculated the depth ($z_T$) where the geotherm reaches the dehydration depth:

$$z_T [\text{m}] = \frac{T_{dehyd} - T_{surf}}{\frac{dT}{dz}}. \quad \text{(A3)}$$

$T_{dehyd}$ is the dehydration temperature of serpentine (550–1000 K; L. Pohl & D. T. Britt 2024), and $T_{surf}$ is the surface temperature (we compare 290 K versus 525 K). For comparison, we also calculated the depth ($z_P$) where the lithostatic pressure reaches approximate dehydration pressures:

$$z_P [\text{m}] = \frac{P_{max,hyd}}{\rho_{crust} \ g}. \quad \text{(A4)}$$

$P_{max,hyd}$ is the pressure where serpentine begins to dehydrate into forsterite and clinoenstatite ($2 \times 10^9$ Pa; S. Liang et al. 2022), $\rho_{crust}$ is the average density of basalt (2900 kg m$^{-3}$), and $g$ [ms$^{-2}$] is the gravitational acceleration. The final dehydration depth ($z_{hyd}$) uses the shallowest depth between the geotherm- and pressure-derived methods.

$$z_{hyd} [\text{m}] = \max(0, \quad \min(z_T, \quad z_P)). \quad \text{(A5)}$$

The maximum hydration of the crust ($\Omega_{H_2O}^{crust}$) is then calculated as

$$\Omega_{H_2O}^{crust} [\text{kg H}_2\text{O}] = A_p \ \rho_{crust} \ z_{hyd} \ \psi_{crust} \quad \text{(A6)}$$

where $A_p$ is the surface area of the planet, and $\psi_{crust}$ is the allowable fraction of hydration in the crust, which we conservatively assume to be 0.5 weight percent (representative of the deep crust) to impose more severe restrictions on ingassing (R. L. Carlson 2003). We used Equations (A1)–(A6) to estimate the maximum crustal water content over time (Figure A3), and found the maximum crustal hydration reservoir to range from ∼0.1%–10% of an Earth ocean.

Motivated by this result, we then repeated our carbon cycle Monte Carlo calculations with a restricted interior water capacity ($\Omega_{H_2O}^{silic} = 10^{18} - 10^{20}$ kg as opposed to our nominal range of $10^{20}$–$10^{22}$ kg). This sensitivity represents a tectonic regime in which only the crust can be hydrated (no subduction of water and no mantle water storage). Figures A4 and A5 show the results of this sensitivity test, alongside the other water ingassing end-member cases described in the main text. Overall, the details of water ingassing do not significantly affect the final temperature of each model run. The initial surface water reservoir is the primary factor controlling the long-term balance of the carbon cycle.

### A.3. Sensitivity of Results to Other Monte Carlo Variables

We explored how all the sampled Monte Carlo variables (Table 2) affected the final surface temperature of the model runs. Figure A6 shows the probability that the final surface ends above 400 K versus nine Monte Carlo parameters, including initial interior water, initial carbon inventories, soil age, etc. Although the initial surface water has the strongest control on the final surface temperature, the initial surface carbon inventory is a secondary control.

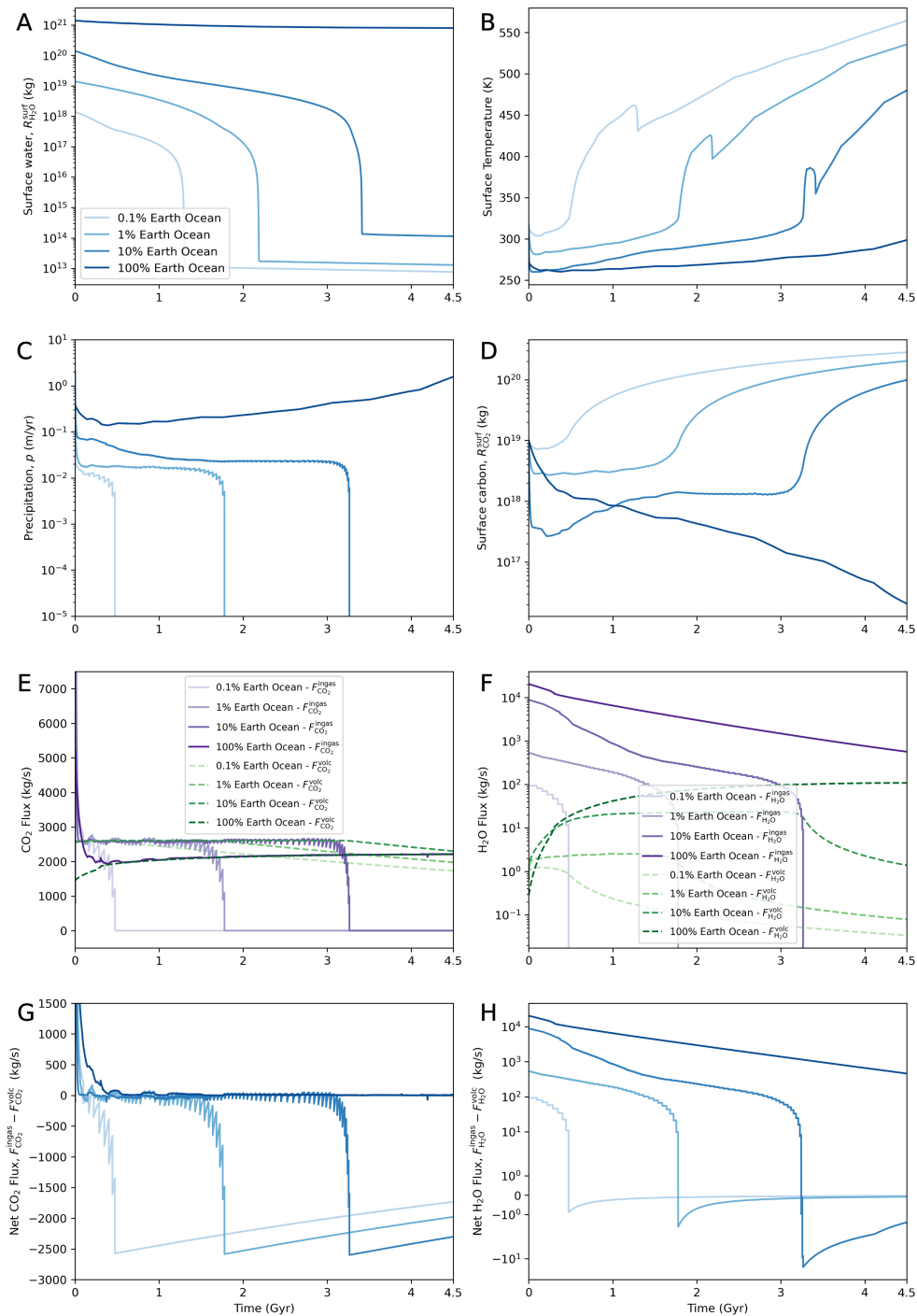

**Figure A1.** Same as Figure 3 except water ingassing is surface dependent rather than mass dependent. Assumes Earth-like hypsometry, instellation, and an albedo ($\alpha$) of 0.3. Higher water inventories are needed to ensure a balanced climate compared to the mass-dependent water ingassing shown in Figure 3.

We measured the timing of runaway warming for Venus-like planets in Figure A7. Models with mass-dependent ingassing take an average of 0.57 Gyr to experience runaway warming, while models with surface-depent ingassing on average take 1.07 Gyr.

### A.4. Sensitivity to Hypsometry, Escape, and Melt Production

We also explored planets with Venus-like instellation but Earth-like hypsometry to further compare the sensitivity of the deep-water cycle to planetary topography (Figure A8). The outcomes of the mass-dependent cases are far more similar compared to the surface-dependent cases. When a planet with Venus-like instellation and surface-dependent deep-water cycling has an Earth-like hypsometry, less water is required to reach the threshold needed for the carbon cycle to be balanced.

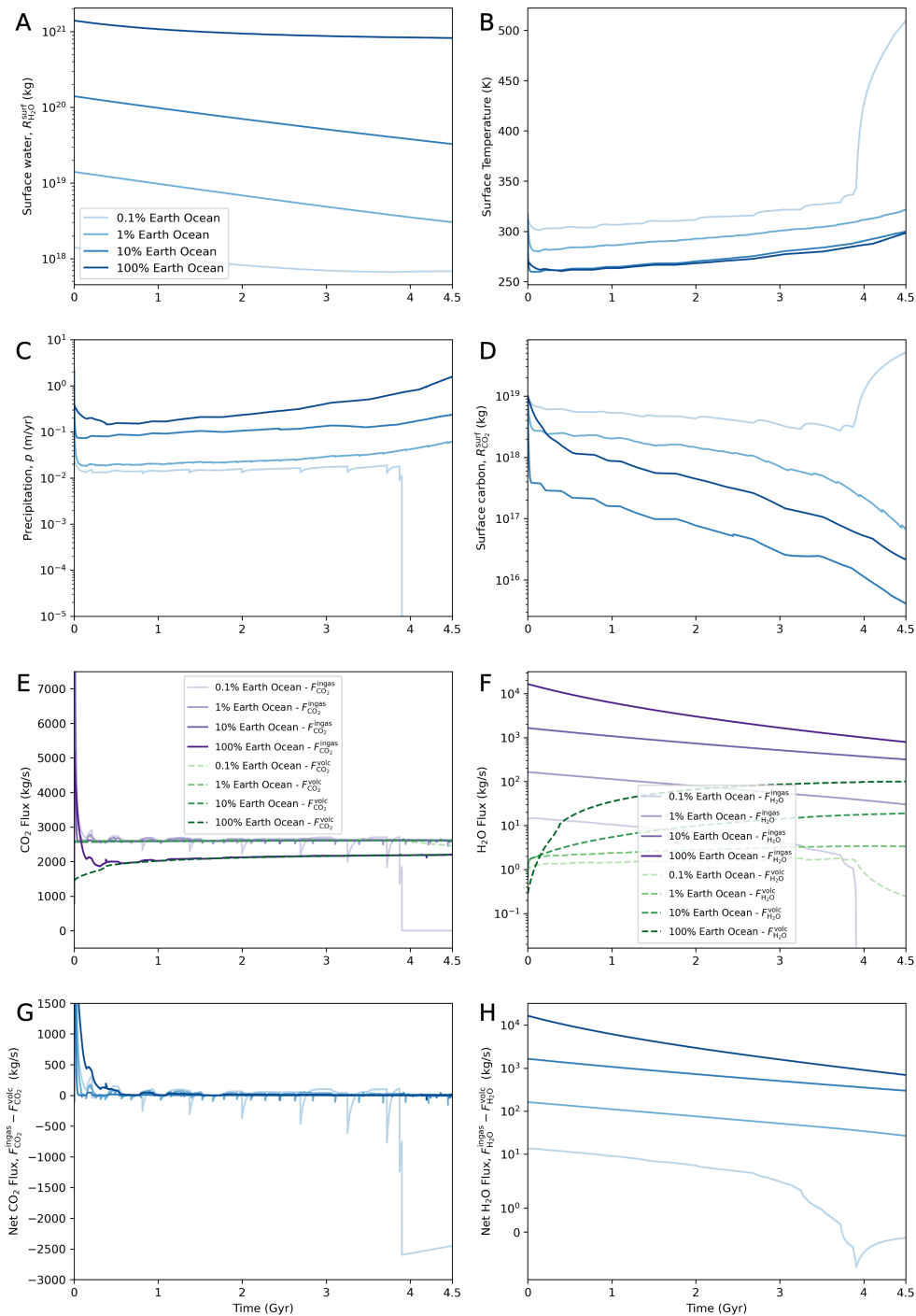

**Figure A2.** Same as Figure 3 except that escape processes are not included. Assumes Earth-like hypsometry, instellation, mass-dependent water ingassing, and an albedo ($\alpha$) of 0.3. Less water is needed to maintain a balanced climate in comparison to outputs that include escape processes (Figure 3) since atmospheric escape enhances water loss.

A supplementary figure, Figure A9, compares models with and without atmospheric escape for mass-dependent ingassing. Overall, the probability that the model runs remain habitable is less affected by the exclusion of atmospheric escape when ingassing is mass dependent. The probabilities are most affected when the surface water reservoir is below ∼1% of an Earth ocean.

All Monte Carlo results presented in the main text use declining melt production over time (described in Section 2.3). We compared different melt production regimes in Figure A10, which includes two constant melt

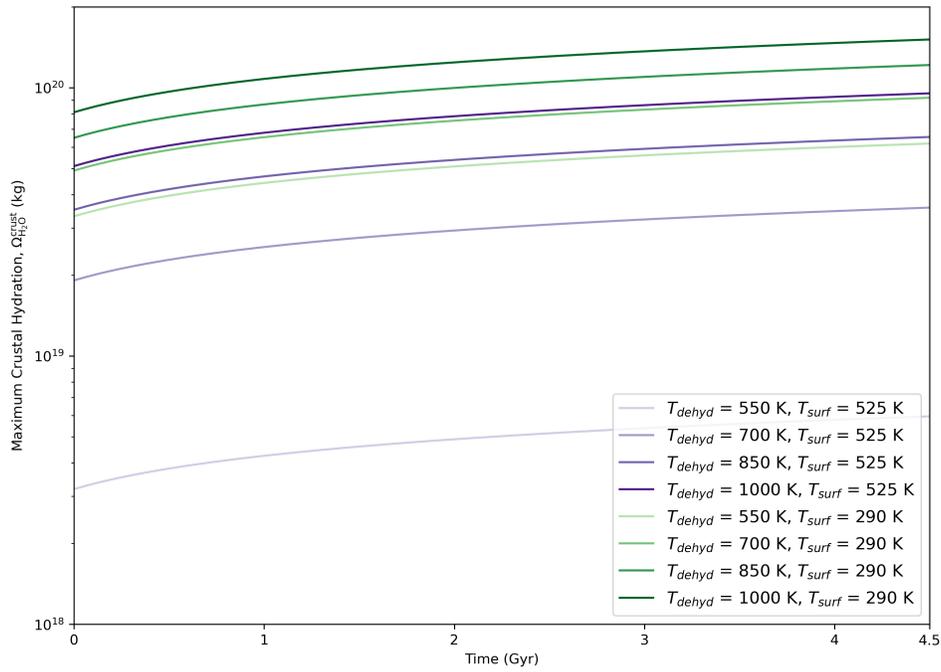

**Figure A3.** Estimates of maximum crustal water inventory from back-of-envelope calculations. The maximum depth is calculated where the geothermal gradient intersects the dehydration temperature of serpentine, and where the lithostatic pressure intersects the dehydration pressure of serpentine (Equations (A1)–(A6)). Crustal hydration is limited to ∼0.1% of an Earth ocean ($10^{18}$ kg), and ∼10% of an Earth ocean ($10^{20}$ kg).

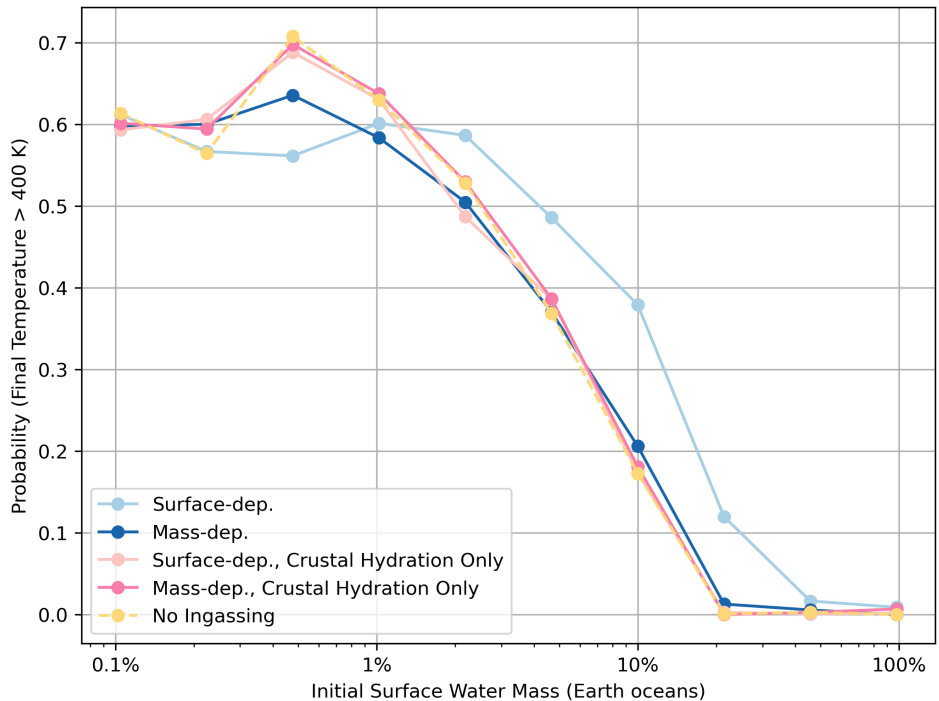

**Figure A4.** Similar to Figure 6, but additionally including sensitivity tests where only the crust can be hydrated ("crustal hydration only," pink lines). Albedo is constant at $\alpha = 0.3$, and escape processes are included for Earth-like insolation and hypsometry. Even with limited or no ingassing, small initial surface reservoirs result in an imbalanced carbon cycle that leads to uninhabitable surfaces.

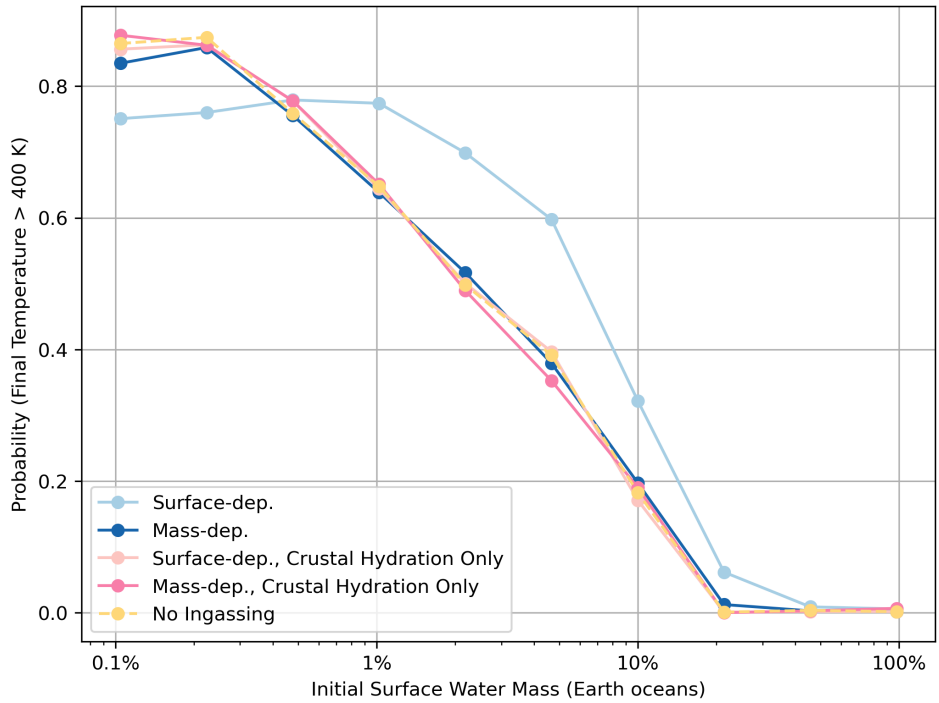

**Figure A5.** Same as Figure A4 except with atmospheric escape processes excluded ($\alpha = 0.3$). Neglecting atmospheric escape processes enhances surface warming for initial water inventories less than 10% of Earth's oceans due to increased greenhouse warming from water vapor. Above 10% of Earth's oceans, the probabilities are similar or better compared to Figure A4, as atmospheric escape may accelerate water loss through H escape.

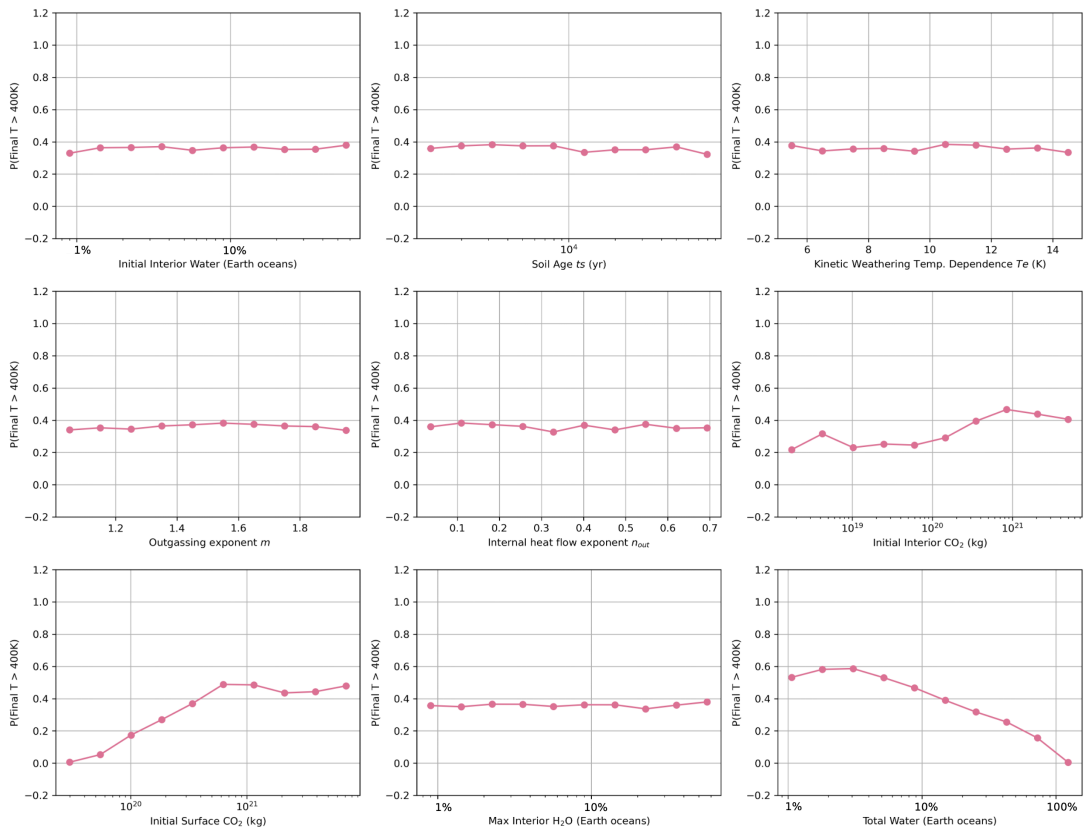

**Figure A6.** The probability of reaching a final surface temperature above 400 K for all unknown Monte Carlo parameters (Table 2). Model outputs for an Earth-like planet (instellation, hypsometry, albedo) with mass-dependent ingassing and atmospheric escape. Besides the initial surface water inventory (Figure 6), the final surface temperature is also dependent on the initial surface carbon inventory, and weakly dependent on the initial interior carbon inventory.

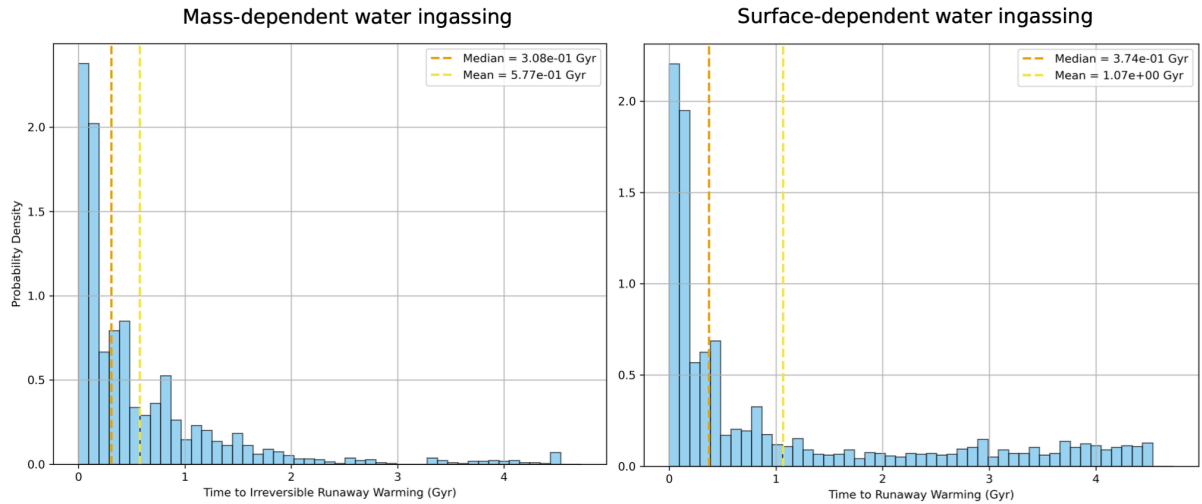

**Figure A7.** Time distribution of when the model runs enter a runaway warming phase for Venus-like instellation, hypsometry, and $\alpha = 0.65$. Mass-dependent water ingassing is shown on the left, and surface-dependent ingassing is shown on the right. Runaway warming is defined as the first instance the model's surface temperature exceeds 400 K and never returns to temperate conditions. The data is divided into 50 equal-width bins (light blue bar plot) where the number of runaway models are divided by the total number of runs to calculate the probability density estimate of when runaway warming is more likely (y-axis). The median and mean are plotted as orange and yellow dotted lines, respectively. Even for optimistic albedo conditions, Venus-like planets that have imbalanced carbon cycles enter the runaway warming phase relatively early, before 2 Gyr.

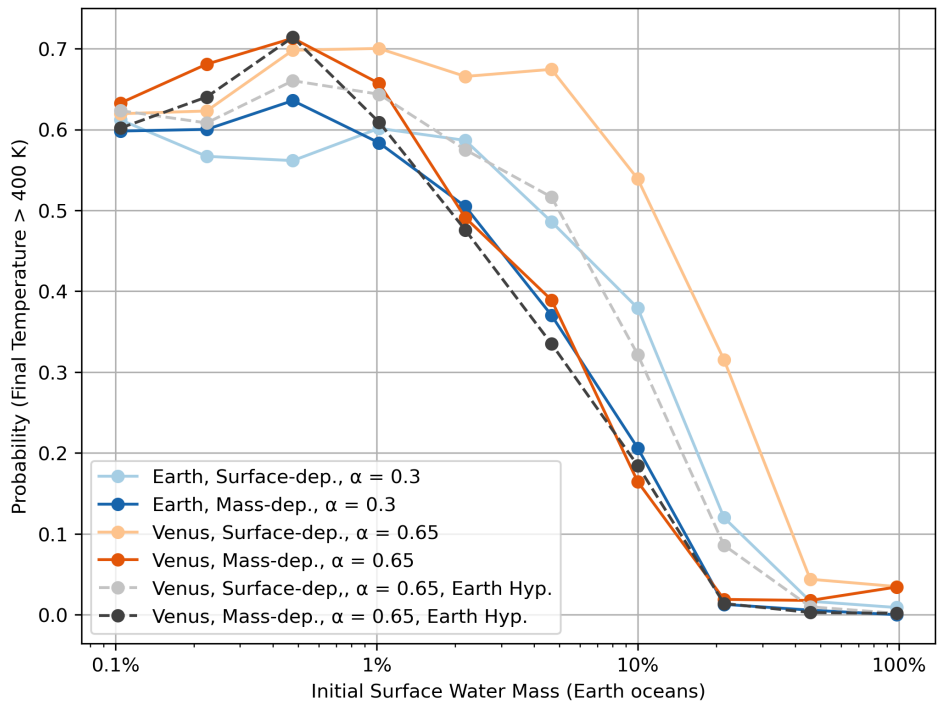

**Figure A8.** Probability plot showing the effect of planetary hypsometry for Earth-like and Venus-like instellations. Higher probabilities imply that the model is more likely to reach temperatures above 400 K. Models with Earth-like hypsometry and instellation are shown in blue shades, and models with Venus-like hypsometry and instellation are shown in orange shades, similar to Figure 7. Models with Earth-like hypsometry but Venus-like instellation are denoted by shades of gray. Lighter shades of any color denote the surface-dependent deep-water cycle.

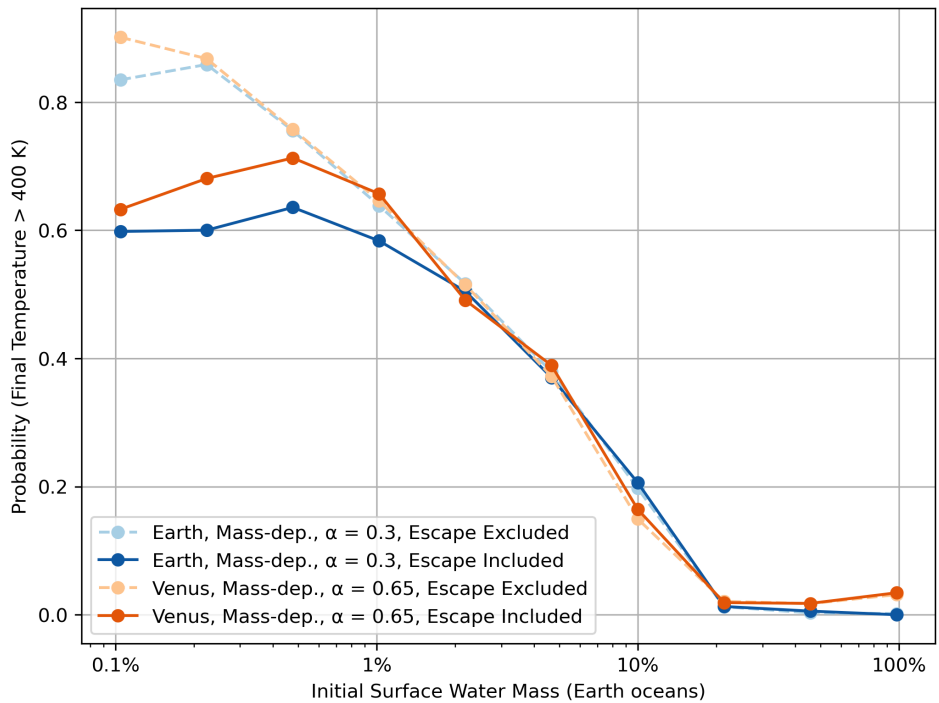

**Figure A9.** Probability plot showing the effect of escape processes at Earth-like and Venus-like instellations. Same as Figure 8 except that all models use mass-dependent water ingassing, as opposed to surface-dependent ingassing. Runs that exclude atmospheric escape are denoted by the dotted lines and lighter shades. Atmospheric escape is less significant to habitability outcomes at higher initial water inventories above ∼1% of Earth's ocean mass.

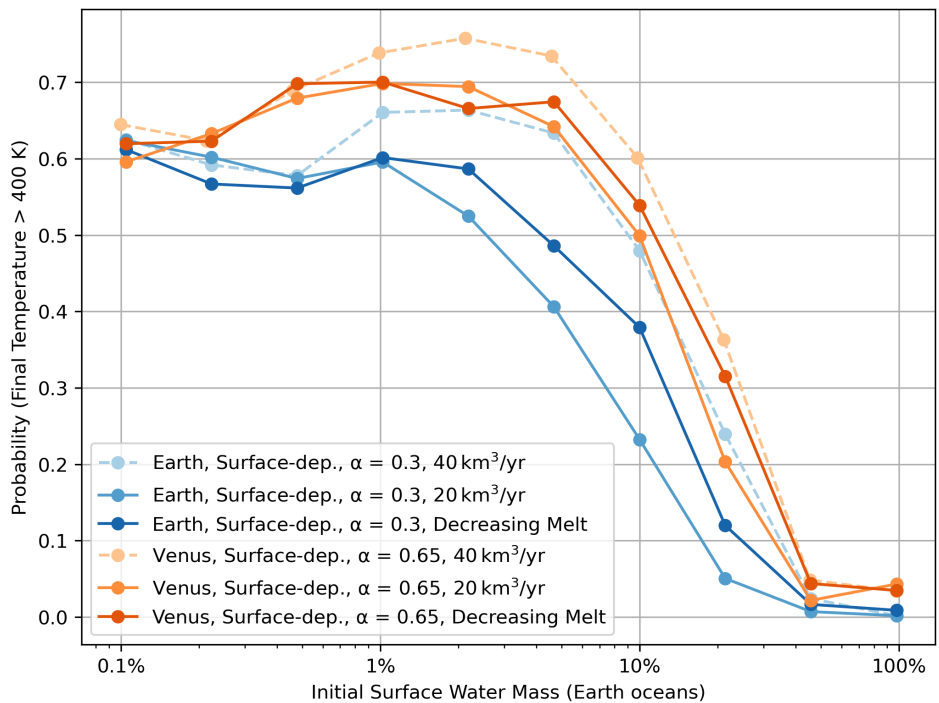

**Figure A10.** Probability plot showing the effect of different melt production regimes for Earth-like and Venus-like instellations. All model runs assume surface-dependent ingassing. Higher probabilities imply that the model is more likely to reach temperatures above 400 K. Earth-like models are in shades of blue, and Venus-like models are in shades of orange, similar to Figure 7. Dotted lines and the lightest shade indicate a constant melt production of 40 km$^3$ yr$^{-1}$. Medium shades represent a constant melt production of 20 km$^3$ yr$^{-1}$. The darkest shades of each color represent the decreasing melt production associated with secular cooling used in all Monte Carlo calculations presented in the main text. Higher constant melt productions require more surface water to maintain habitable surface temperatures.

production regimes (20 km$^3$ yr$^{-1}$ and 40 km$^3$ yr$^{-1}$) and the nominal declining melt production regime. A constant high melt production rate requires more surface water to maintain temperate surface conditions compared to low constant melt production and decreasing melt production.

ORCID iDs

Haskelle T. White-Gianella  https://orcid.org/0009-0004-7751-8964
Joshua Krissansen-Totton  https://orcid.org/0000-0001-6878-4866